\documentclass[aps,twocolumn]{revtex4-1}
\usepackage{amsmath,amssymb}
\usepackage{amsthm}
\usepackage{accents}
\usepackage{nicefrac}
\usepackage[ruled,vlined]{algorithm2e}
\usepackage{graphicx}

\usepackage[T1]{fontenc}
\usepackage[latin9]{inputenc}
\usepackage{color}
\usepackage[english]{babel}

\begin{document}
\title{Projected and Hidden Markov Models for calculating kinetics and metastable states of complex molecules}
  
\author{Frank No\'{e}}
\email{frank.noe@fu-berlin.de}
\thanks{corresponding author}
\affiliation{Department of Mathematics and Computer Science, FU Berlin, Arnimallee 6, 14159 Berlin,
Germany}
  
\author{Hao Wu}
\thanks{equal contribution}
\affiliation{Department of Mathematics and Computer Science, FU Berlin, Arnimallee 6, 14159 Berlin,
Germany}

\author{Jan-Hendrik Prinz},
\thanks{equal contribution}
\affiliation{Department of Mathematics and Computer Science, FU Berlin, Arnimallee 6, 14159 Berlin,
Germany}

\author{Nuria Plattner},
\affiliation{Department of Mathematics and Computer Science, FU Berlin, Arnimallee 6, 14159 Berlin,
Germany}

\begin{abstract}
Markov state models (MSMs) have been successful in computing metastable states, slow relaxation timescales and associated structural changes, and stationary or kinetic experimental observables of complex molecules from large amounts of molecular dynamics simulation data. However, MSMs approximate the true dynamics by assuming a Markov chain on a clusters discretization of the state space. This approximation is difficult to make for high-dimensional biomolecular systems, and the quality and reproducibility of MSMs has therefore been limited. Here, we discard the assumption that dynamics are Markovian on the discrete clusters. Instead, we only assume that the full phase-space molecular dynamics is Markovian, and a projection of this full dynamics is observed on the discrete states, leading to the concept of Projected Markov Models (PMMs). Robust estimation methods for PMMs are not yet available, but we derive a practically feasible approximation via Hidden Markov Models (HMMs). It is shown how various molecular observables of interest that are often computed from MSMs can be computed from HMMs / PMMs. The new framework is applicable to both, simulation and single-molecule experimental data. We demonstrate its versatility by applications to educative model systems, an 1 ms Anton MD simulation of the BPTI protein, and an optical tweezer force probe trajectory of an RNA hairpin.
\end{abstract}

\maketitle

Conformational transitions are essential to the function of proteins
and nucleic acids. With the ever increasing time resolution of ensemble
kinetics experiments and the more recent maturation of sensitive single-molecule
techniques in biophysics, experimental evidence supporting the near-universality
of the existence of multiple metastable conformational substates and
complex kinetics in biomolecules has continued to accumulate \cite{GansenSeidel_PNAS2009_Nucleosome,NeubauerSeidel_JACS2007_Rhodamine,Xie_PRL05_PowerLaw,EisenmesserKayKern_Nature2005,Santoso_PNAS2009_FretPolymerase,GebhardRief_PNAS10_AFMEnergyLandscapeProtein,WensleyClark_Nature09_FrustrationHelix}.
Markov (state) models (MSMs) are a very successful approach to deal
with such a multitude of metastable states, that has emerged from
the simulation community \cite{SchuetteFischerHuisingaDeuflhard_JCompPhys151_146,SwopeEtAl_JPCB108_6582,SinghalPande_JCP123_204909,NoeHorenkeSchutteSmith_JCP07_Metastability,ChoderaEtAl_JCP07,BucheteHummer_JPCB08,NoeSchuetteReichWeikl_PNAS09_TPT}.
A MSM consists of a discretization of the molecular state space into
$n$ clusters, and a $n\times n$ transition probability matrix containing
the conditional probabilities that the system will, given that it
is in one of its $n$ discrete substates, be found in any of these
$n$ discrete substates a fixed lag time $\tau$ later. Because only
\emph{conditional} transition probabilities are needed, an MSM can
be estimated from ensembles of short trajectories, computed distributedly
on clusters or volunteer networks \cite{NoeSchuetteReichWeikl_PNAS09_TPT,VoelzPande_JACS10_NTL9,BowmanGeissler_PNAS2012_BindingSiteMSM}.
This circumvents the need for ultralong trajectories that can only
be computed by special-purpose supercomputers \cite{Shaw_Science10_Anton,LindorffLarsenEtAl_Science11_AntonFolding}.
Additionally, MSMs have been so successful because they permit many
important thermodynamic, kinetic and mechanistic molecular quantities
to be computed much more directly and unambiguously than with conventional
MD analyses.

However, a key approximation of MSMs is that they assume a Markov
chain on the discrete clusters --- although these discrete dynamics
are not Markovian. It has been rigorously shown that the MSM approximation
can be very precise if the molecular coordinates relevant for the
slow transitions are finely discretized \cite{SarichNoeSchuette_MMS09_MSMerror,PrinzEtAl_JCP10_MSM1}.
In practice the discretization quality will depend on the subset molecular
coordinates and metric used as input, and the method used to cluster
this coordinate space. The sheer high-dimensionality of solvated biomolecular
systems, and the necessity to neglect many coordinates (velocities,
solvent positions), limits the practical ability to produce a very
fine discretization. Therefore, MSM results may significantly differ
depending on the choice of input coordinates and clustering methods
\cite{SchwantesPande_JCTC13_TICA,PerezEtAl_JCP13_TICA,PrinzEtAl_JCP10_MSM1,BucheteHummer_JPCB08,ChoderaEtAl_JCP07}.
Moreover, the assumption that the dynamics of the clustered molecular
observables is Markovian prohibits the use of MSMs for the analysis
of experimental single-molecule trajectories, where the molecular
coordinate traced is determined by what is experimentally observable
and cannot be arbitrarily chosen.

Here, we introduce a new framework that altogether discards the assumption
that dynamics are Markovian on the observed clusters. Instead we only
make very basic physical assumptions: The full phase-space dynamics
are Markovian, and in thermodynamic equilibrium. This full-space dynamics
becomes projected onto the discrete clusters whose discrete dynamics
is observed. This leads to the concept of Projected Markov Models
(PMMs). We show that if the dynamics are metastable, having a number
$m$ slow relaxation processes, and if there is a separation of timescales
to the next-faster relaxation processes, then PMMs can be approximated
by Hidden Markov Models (HMMs) with $m$ hidden states. We describe
an MSM$\rightarrow$HMM transformation that provides a good starting
point to estimate the HMM via Baum-Welch Expectation Maximization
algorithm. It is shown how various molecular observables of interest
that are often computed from MSMs can be computed from HMMs. The new
method is applicable to both, simulation and single-molecule experimental
data. Moreover, all important thermodynamic, kinetic and mechanistic
molecular quantities computable from MSMs can also be computed from
HMMs. We demonstrate the versatility of our approach by applications
to various systems --- including model systems that demonstrate the
superiority of PMM/HMM models over MSMs, an 1 ms Anton MD simulation
of the BPTI protein where a three-state rate matrix with metastable
sets of structures is readily obtained, and an optical tweezer force
probe trajectory of an RNA hairpin, where a hidden and yet unreported
transition state is found.

\section{Projected Markov Models}

We assume that there is a Markov process $\{\mathbf{z}_{t}\}$ in
state space $\Omega$ (a state may consist of both positions and velocities
- depending on the model of the dynamics). This Markov process is
assumed to be ergodic and reversible with respect to a unique stationary
distribution $\mu(\mathbf{z})$. Often, a canonical ensemble is employed,
and then the stationary distribution is the Boltzmann distribution
$\mu(\mathbf{z})=Z^{-1}\mathrm{e}^{-\beta H(\mathbf{z})}$ with $H(\mathbf{z})$
the total energy and $\beta=1/k_{B}T$ the inverse temperature. For
such a process, we can write the ensemble dynamics as follows: We
consider a probability distribution of states $p_{0}$. At a later
time $\tau$, the distribution will have evolved according to the
Markov propagator $\mathcal{P}$:
\[
p_{\tau}(\mathbf{z}_{\tau})=\mathcal{P}(\tau)\: p_{0}(\mathbf{z}_{0})=\int_{\mathbf{z}_{0}}p_{\tau}(\mathbf{z}_{0},\mathbf{z}_{\tau})\: p_{0}(\mathbf{z}_{0}).
\]
where the conditional transition probability $p_{\tau}(\mathbf{z}_{0},\mathbf{z}_{\tau})$
characterizes the dynamics of the system. With the ergodicity, we
can expand the propagation density into basis functions 
\begin{eqnarray*}
p_{\tau}(\mathbf{z}_{0},\mathbf{z}_{\tau}) & = & \mu(\mathbf{z}_{\tau})+\sum_{i=2}^{\infty}\mathrm{e}^{-\kappa_{i}\tau}\frac{\phi_{i}(\mathbf{z}_{0})}{\mu(\mathbf{z}_{0})}\phi_{i}(\mathbf{z}_{\tau})
\end{eqnarray*}
where $\kappa_{i}$ is the relaxation rate of the $i$th-slowest process
and $t_{i}=\kappa_{i}^{-1}$ is the corresponding relaxation timescale.
We can also consider the corresponding correlation density, i.e. the
joint probability density to observe the system at position $\mathbf{z}_{0}$
at time $0$ and at position $\mathbf{z}_{\tau}$ at time $\tau$:
\begin{eqnarray*}
c_{\tau}(\mathbf{z}_{0},\mathbf{z}_{\tau}) & = & \mu(\mathbf{z}_{0})\mu(\mathbf{z}_{\tau})+\sum_{i=2}^{\infty}\mathrm{e}^{-\kappa_{i}\tau}\phi_{i}(\mathbf{z}_{0})\phi_{i}(\mathbf{z}_{\tau})
\end{eqnarray*}
Note that for $\tau\rightarrow\infty$, the joint probability density
is simply given by the stationary probabilities: $c_{\infty}(\mathbf{z}_{0},\mathbf{z}_{\infty})=\mu(\mathbf{z}_{0})\mu(\mathbf{z}_{\infty})$.
For the rest of the paper we assume that our system of interest has
$m$ slow processes and a timescale separation to the faster processes.
Thus, at lag times significantly larger than $t_{m}=\kappa_{m}^{-1}$,
the correlation density is approximately given by: 
\begin{eqnarray}
c_{\tau}(\mathbf{z}_{0},\mathbf{z}_{\tau}) & \approx & \mu(\mathbf{z}_{0})\mu(\mathbf{z}_{\tau})+\sum_{i=2}^{m}\mathrm{e}^{-\kappa_{i}\tau}\phi_{i}(\mathbf{z}_{0})\phi_{i}(\mathbf{z}_{\tau})\label{eq_correlation-density-m}
\end{eqnarray}
Now we assume that the molecular state space (typically only configurations,
not velocities) is completely partitioned into a set of $n$ clusters
$\{S_{i}\}$, which might be rather coarse. What happens to the dynamics
when we observe it on the space of clusters? From Eq. (\ref{eq_correlation-density-m})
we can compute the correlation matrix between clusters:
\begin{eqnarray*}
C_{ij} & = & \int_{\mathbf{z}_{0}\in S_{i}}d\mathbf{z}_{0}\int_{\mathbf{z}_{\tau}\in S_{j}}d\mathbf{z}_{\tau}\: c_{\tau}(\mathbf{z}_{0},\mathbf{z}_{\tau})\\
 & = & \pi_{i}\pi_{j}+\sum_{k=2}^{m}\mathrm{e}^{-\kappa_{k}\tau}q_{ki}q_{kj}
\end{eqnarray*}
where $\pi_{i}=\int_{\mathbf{z}\in S_{i}}d\mathbf{z}\:\mu(\mathbf{z})$
are the stationary probabilities and $\mathbf{q}_{k}$ is the $k$th
discretized eigenfunction:
\[
q_{ki}=\int_{\mathbf{z}_{0}\in S_{i}}d\mathbf{z}_{0}\:\phi_{k}(\mathbf{z}_{0}).
\]
We can also express the correlation matrix as
\begin{equation}
\mathbf{C}=\mathbf{Q}^{\top}\widetilde{\boldsymbol{\Lambda}}\mathbf{Q}\label{eq_PMM_C}
\end{equation}
where $\mathbf{Q}\in\mathbb{R}^{m\times n}$ contains the discretized
projected eigenfunctions $\mathbf{q}_{k}$, and $\widetilde{\boldsymbol{\Lambda}}\in\mathbb{R}^{m\times m}$
contains the $m$ dominant eigenvalues. We will use the tilde in order
to annotate ``small'' matrices or vectors related to the $m$ metastable
processes. If we write the stationary probability vector $\boldsymbol{\pi}$
on the diagonal of the matrix $\boldsymbol{\Pi}$, we can write the
transition matrix between clusters as:
\begin{equation}
\mathbf{T}(\tau)=\boldsymbol{\Pi}^{-1}\mathbf{Q}^{\top}\widetilde{\boldsymbol{\Lambda}}(\tau)\mathbf{Q}\label{eq_PMM_T}
\end{equation}
This is the transition matrix that is estimated when building a Markov
model at lag time $\tau$. Now we can easily illustrate the problem
of MSMs: The dynamics between clusters are not Markovian, i.e. the
transition matrix estimated at $\tau$ cannot be used to predict long-timescale
behavior:
\begin{eqnarray*}
\mathbf{T}(2\tau) & = & \boldsymbol{\Pi}^{-1}\mathbf{Q}\widetilde{\boldsymbol{\Lambda}}(2\tau)\mathbf{Q}^{T}=\boldsymbol{\Pi}^{-1}\mathbf{Q}[\widetilde{\boldsymbol{\Lambda}}(\tau)]^{2}\mathbf{Q}^{T}\\
 & \neq & \boldsymbol{\Pi}^{-1}\mathbf{Q}\widetilde{\boldsymbol{\Lambda}}(\tau)\mathbf{Q}^{T}\boldsymbol{\Pi}^{-1}\mathbf{Q}\widetilde{\boldsymbol{\Lambda}}(\tau)\mathbf{Q}^{T}=\mathbf{T}^{2}(\tau)
\end{eqnarray*}
The first row is not equal to the second row because projected eigenvectors
$\mathbf{Q}$ of the full-space dynamics are \emph{not} eigenvectors
of $\mathbf{T}$, and are therefore \emph{not} orthonormal with respect
to the observed stationary distribution: $\mathbf{Q}^{T}\boldsymbol{\Pi}^{-1}\mathbf{Q}\neq\mathbf{Id}$.
Therefore, in order to estimate the cluster dynamics in a way that
is unbiased, and that allows the long-time dynamics to be predicted,
one needs to estimate the PMM quantities
\[
\{\mathbf{Q},\widetilde{\boldsymbol{\Lambda}}\}
\]
separately.

\section{Approximating PMMs via Hidden Markov Models}

\label{sec_2}

In general, estimating the matrices $\{\mathbf{Q},\widetilde{\boldsymbol{\Lambda}}\}$
is difficult, especially for large $m$ and $n$. Therefore we consider
a slightly different model that we can efficiently estimate: A hidden
Markov model (HMM). A hidden Markov model consists of a transition
matrix between $m$ hidden (here metastable) states, $\widetilde{\mathbf{T}}(\tau)$,
and an associated stationary distribution $\widetilde{\boldsymbol{\pi}}$.
This hidden dynamics has a joint probability (correlation) matrix
$\widetilde{\mathbf{C}}=\widetilde{\boldsymbol{\Pi}}\widetilde{\mathbf{T}}$.
Each hidden state $i$ will output to one of the observable states
$j$ with a probability $\chi_{ij}$, such that the vector $\boldsymbol{\chi}_{i}$
is the output probability distribution of hidden state $i$. We can
write the correlation matrix on the observed states as:

\begin{eqnarray}
\mathbf{C} & = & \boldsymbol{\chi}^{\top}\widetilde{\mathbf{C}}\boldsymbol{\chi}=\boldsymbol{\chi}^{\top}\widetilde{\boldsymbol{\Pi}}\widetilde{\mathbf{T}}\boldsymbol{\chi}\label{eq_HMM1}\\
 & = & \boldsymbol{\chi}^{\top}\widetilde{\mathbf{L}}^{\top}\widetilde{\boldsymbol{\Lambda}}\widetilde{\mathbf{L}}\boldsymbol{\chi}\nonumber \\
 & = & \:\:\:\:\mathbf{Q}^{\top}\widetilde{\boldsymbol{\Lambda}}\mathbf{Q}\nonumber 
\end{eqnarray}

By comparing the last row to (\ref{eq_PMM_C}), it is apparent that
a HMM has a similar structure like a PMM. Here, the vectors in $\mathbf{Q}$
are given by the HMM eigenvectors $\widetilde{\mathbf{L}}$ projected
onto the observable states via $\boldsymbol{\chi}$. However, we want
to use HMM estimation algorithms to estimate the slow molecular kinetics
of a Markov process observed on a cluster space, which is a PMM (\ref{eq_PMM_C})
--- and therefore we must show that a PMM can also be represented
as a HMM. This is not obvious: A given PMM, defined by the slow process
eigenfunctions and the chosen discretization has a certain $\mathbf{Q}$.
It is not a priori clear whether this $\mathbf{Q}$-matrix can be
represented by decomposition into the two matrices $\boldsymbol{\chi}\widetilde{\mathbf{L}}$,
because these matrices have to fulfill the constraints that the columns
of $\boldsymbol{\chi}$ are probability distributions and the rows
of $\widetilde{\mathbf{L}}$ form a set of eigenvectors which are
orthonormal with respect to $\widetilde{\boldsymbol{\pi}}^{-1}$.
Appendix A contains a proof that modeling a PMM with an HMM is valid
in a special, but interesting case. We summarize it as follows:

\emph{Given a Markov process $\{\mathbf{x}_{t}\}$ that is ergodic
and reversible with respect to the unique stationary distribution
$\mu(\mathbf{x})$. Given that this process has $m$ metastable states,
such that there is a gap in the relaxation timescales, $t_{m+1}\ll t_{m}$,
and the stationary distribution $\mu(\mathbf{x})$ almost decomposes
into $m$ modes, such that almost all stationary probability is in
the metastable states and the intervening transition states have vanishing
populations. We further consider an arbitrary discretization of the
state space $\mathbf{x}$ into $n$ clusters. Then, the dynamics on
the $n$ discrete states is described by a discrete hidden Markov
model with $m$ hidden and $n$ observed states.}

This is an important result: in many applications, especially in biomolecular
dynamics, we have a few metastable states with rarely populated transition
regions. The theorem above says, that even using a \emph{poor} discretization
of the state space of such a system, we can still describe its metastable
dynamics \emph{exactly} with an HMM. Of course, our practical ability
to find the true HMM will depend on the amount of statistics at hand,
and may very well depend on the quality of the discretization. However,
we will show in the application section that HMMs perform very well
in this setting, and almost exclusively better than MSMs.

\subsection{Initializing a hidden Markov model from a Markov model}

Estimating hidden Markov models is more difficult than estimating
directly observed Markov models, because in constast to the MSM likelihood,
the HMM likelihood does not necessarily have a unique optimum. Therefore,
it is important to start the HMM estimation ``close'' to the optimal
result. How do we get a good initial guess for the hidden transition
matrix $\widetilde{\mathbf{T}}(\tau)$ and the output probability
matrix $\boldsymbol{\chi}$?

Hence we propose an initial HMM based on a direct Markov model. Given
the simulation trajectories, discretized in $n$ states, we estimate
a Markov model transition matrix at some lag time $\tau$, $\mathbf{T}(\tau)\in\mathbb{R}^{n\times n}$.
In order to ensure that this Matrix fulfills detailed balance, we
use the reversible transition matrix estimator described in \cite{PrinzEtAl_JCP10_MSM1}
and implemented in EMMA \cite{SenneSchuetteNoe_JCTC12_EMMA1.2}.

Next, we fix a number of hidden states, $m$, and obtain an initial
estimate of the output probability matrix $\boldsymbol{\chi}$. For
this, we first employ the PCCA+ method \cite{DeuflhardWeber_PCCA}
implemented in EMMA \cite{SenneSchuetteNoe_JCTC12_EMMA1.2}. PCCA+
provides, for each observed state $i$ a degree of membership to a
metastable state $j$, $m_{ij}$. PCCA+ does this by first proposing
$m$ observed states as representatives of the metastable states,
each obtaining membership 1 to the respective metastable states and
0 to the others. This is an approximation that will later be lifted
by the HMM optimization. The full membership matrix $\mathbf{M}\in\mathbb{R}^{n\times m}$
is obtained by solving a linear system of equations, as described
in \cite{DeuflhardWeber_PCCA}. The membership matrix has the property
that its rows sum up to 1, but the implementation described in \cite{DeuflhardWeber_PCCA}
has the undesirable property that membership values could be negative.
We currently avoid this by setting negative memberships to 0 and then
renormalizing all rows of $\mathbf{M}$. This intervention is avoided
by PCCA++ \cite{Roeblitz_PhD} which generates non-negative membership
matrices and will be used in the future.

Now, the membership $m_{ij}$ can be interpreted as a probability
of being in a metastable state $j$, given that the system is observed
in discrete state $i$. We can use the membership matrix to coarse-grain
the stationary probabilities to the hidden metastable states:
\begin{eqnarray*}
\widetilde{\boldsymbol{\pi}} & = & \mathbf{M}^{\top}\boldsymbol{\pi}
\end{eqnarray*}
and we can use Bayesian statistics in order to transform $\mathbf{M}$
to the desired output probabilities:
\begin{eqnarray*}
\mathbb{P}[\mathrm{cluster}\: j\mid\mathrm{hidden}\: i] & = & \frac{\mathbb{P}[\mathrm{cluster}\: j]}{\mathbb{P}[\mathrm{hidden}\: i]}\mathbb{P}[\mathrm{hidden}\: i\mid\mathrm{cluster}\: j]\\
\chi_{ij} & = & \frac{\pi_{j}}{\widetilde{\pi}_{i}}m_{ji}.
\end{eqnarray*}
In matrix form:
\begin{equation}
\boldsymbol{\chi}=\widetilde{\boldsymbol{\Pi}}^{-1}\mathbf{M}^{\top}\boldsymbol{\Pi}.\label{eq_HMM-Chiinit}
\end{equation}
Finally, we need a hidden transition matrix $\widetilde{\mathbf{T}}$
which fulfills Eq. (\ref{eq_HMM1}), i.e. which produces, together
with $\boldsymbol{\chi}$, the observable correlation matrix $\mathbf{C}$.
A method for computing such a matrix is given in \cite{KubeWeber_JCP07_CoarseGraining}.
Using their Eq. (12) and performing some algebraic transformations
given in Appendix B.1, we obtain the result:
\begin{equation}
\widetilde{\mathbf{T}}=\mathbf{M}^{\top}\mathbf{T}\mathbf{M}(\mathbf{M}^{\top}\mathbf{M})^{-1}\label{eq_HMM-Tinit}
\end{equation}
which has a nice interpretation: $\mathbf{M}^{\top}\mathbf{T}\mathbf{M}$
performs a coarse-graining of $\mathbf{T}$, and $\mathbf{M}^{\top}\mathbf{M}$
is an overlap matrix needed for normalization. $\widetilde{\mathbf{T}}$
has the nice property that it preserves the dominant kinetics of $\mathbf{T}$:
the eigenvalues of $\widetilde{\mathbf{T}}$ are identical to the
dominant $m$ eigenvalues of $\mathbf{T}$. In some cases $\widetilde{\mathbf{T}}$
may have (usually only slightly) negative elements. Moreover, for
numerical reasons in computing Eq. (\ref{eq_HMM-Tinit}), $\widetilde{\mathbf{T}}$
may no longer exactly fulfill detailed balance. We correct for this
by computing the correlation matrix, $\widetilde{\mathbf{C}}=\widetilde{\boldsymbol{\Pi}}\widetilde{\mathbf{T}}$,
symmetrizing it $\widetilde{\mathbf{C}}\leftarrow\nicefrac{1}{2}(\widetilde{\mathbf{C}}+\widetilde{\mathbf{C}}^{\top})$,
setting negative elements to 0, and then renormalizing the rows of
$\widetilde{\mathbf{C}}$ to obtain $\widetilde{\mathbf{T}}$. Note
that this intervention is important to make sure that the HMM optimization
is seeded with a $\widetilde{\mathbf{T}}$ matrix that has a meaningful
structure, but should not strongly affect the results as $\widetilde{\mathbf{T}}$
will be subsequently optimized.

\subsection{Hidden Markov model estimation}

Consider the observed trajectory $\{s_{t}\}$ and hidden trajectory
$\{h_{t}\}$. The HMM likelihood is given by:
\begin{equation}
\mathbb{P}(\{s_{t}\}\mid\tilde{\mathbf{T}},\boldsymbol{\chi})=\prod_{\begin{array}{c}
\mathrm{all}\:\mathrm{hidden}\:\mathrm{paths}\\
h_{0},...,h_{t_{max}}
\end{array}}\widetilde{\pi}_{h_{0}}\:\chi_{s_{0}h_{0}}\prod_{t=1}^{t_{max}}\widetilde{T}_{h_{t-1}h_{t}}\:\chi_{s_{t}h_{t}}\label{eq_HMM-likelihood}
\end{equation}
Obviously, the product over all possible hidden paths cannot be directly
computed due to a combinatorial explosion of possibilities. The likelihood
(\ref{eq_HMM-likelihood}) can be maximized by a Expectation-Maximization
algorithm, more precisely by the Baum-Welch method \cite{BaumEtAl_AnnMathStatist70_BaumWelch,Welch_IEEE03_BaumWelchImplementation}.
See \cite{Rabiner_IEEE89_HMM} for a thorough and educative description
of HMMs and the Baum-Welch method. 

Since the EM method is established, we give a brief summary of our
implemention in the appendix. EM iterates two steps, called expectation
and maximization step. While the expectation step is general, the
maximization step must be designed for the specific HMM implementation.
Here, we use the Baum-Welch algorithm to estimate a count matrix $\widetilde{\mathbf{Z}}(\mbox{\ensuremath{\tau}})$
containing the estimated numbers of transitions between the $m$ hidden
states, and then estimate the maximum likelihood transition matrix
$\widetilde{\mathbf{T}}(\mbox{\ensuremath{\tau}})$ that fulfills
detailed balance using the algorithm described in \cite{PrinzEtAl_JCP10_MSM1}
and implemented in EMMA \cite{SenneSchuetteNoe_JCTC12_EMMA1.2}. The
HMM is assumed to be in equilibrium, i.e. it uses the stationary probability
distribution of $\widetilde{\mathbf{T}}(\mbox{\ensuremath{\tau}})$
as an initial distribution. The output probabilities $\boldsymbol{\chi}$
are estimated through straightforward histograms of the expected counts
on the clusters

\subsection{Implied timescale plot }

A commonly used approach to assess the quality of a MSM introduced
by \cite{SwopeEtAl_JPCB108_6582} is the implied timescale plot. Here,
one asks how much the dynamics on the discretized state space deviates
from a Markov chain. For a Markov chain, the Chapman-Kolmogorow equality
$[\mathbf{T}(\tau_{0})]^{n}=\mathbf{T}(n\tau_{0})$ holds, and therefore
for every eigenvalue $[\lambda_{i}(\tau_{0})]^{n}=[\lambda_{i}(n\tau_{0})]$.
This condition is equivalent to the condition that the relaxation
timescales (or implied timescales) 
\begin{equation}
t_{i}(\tau)=-\frac{\tau}{\ln|\lambda_{i}(\tau)|}\label{eq_implied-timescale}
\end{equation}
are constant in $\tau=n\tau_{0}$. Because the dynamics on the discretized
state space are not Markovian, the timescales (\ref{eq_implied-timescale})
are \emph{not} constant in $\tau$. In the limit of good statistics
they are guaranteed to be smaller than the true relaxation timescales
\cite{DjurdjevacSarichSchuette_MMS10_EigenvalueError,NoeNueske_MMS13_VariationalApproach},
and the error between the estimated relaxation timescale $t_{i}(\tau)$
and the true relaxation timescale decays slowly, as $\tau^{-1}$ \cite{PrinzChoderaNoe_PRL11_RateTheory}.

Here, we also conduct implied timescale plots in order to get a first
assessment of the quality and robustness of the PMM estimation. However,
instead of computing $\lambda_{i}(\tau)$ from a diagonalization of
the transition matrix on the discretized state space, we use the eigenvalues
of the hidden transition matrix, i.e. the timescales: 
\begin{equation}
\tilde{t}_{i}(\tau)=-\frac{\tau}{\ln|\tilde{\lambda}_{i}(\tau)|}\label{eq_implied-timescale-1}
\end{equation}

If we are in a setting valid for PMM's, i.e. $\tau\gg t_{m+1}$ (all
timescales that are not resolved by the PMM have decayed, where '$\gg$'
is already given by a factor of 2-3), and we are in the limit of good
statistics, then the PMM/HMM estimate of $\tilde{t}_{i}(\tau)$ should
indeed be constant in $\tau$.

\subsection{Hidden Markov Model validation}

Finally, we estimate the HMM at a lag time $\tau_{0}$ that has been
selected such that the relaxation timescales $\tilde{t}_{i}(\tau)$
are constant at lag times $\tau>\tau_{0}$ or larger. Validation of
the model consists of using it to compute kinetic quantities at a
series of lag times $\tau$ and comparing them with the directly computed
quantities at these different lag times. An example is the set-based
Chapman-Kolmogorow test suggested in \cite{PrinzEtAl_JCP10_MSM1}.
Here, we suggest a very simple and direct test based on the relaxation
timescales. Given the HMM estimated at $\tau_{0}$, we compute the
predicted transition matrices for the discretized state space at lag
times $\tau=n\tau_{0}$:
\[
\mathbf{T}^{\mathrm{pred}}(\tau)=\boldsymbol{\Pi}^{-1}\boldsymbol{\chi}^{\top}\widetilde{\boldsymbol{\Pi}}[\widetilde{\mathbf{T}}(\tau_{0})]^{n}\boldsymbol{\chi}
\]
and compare their relaxation timescales with the relaxation timescales
computed directly from MSM transition matrices estimated at $\tau$:
\[
t_{i}^{\mathrm{pred}}(\tau)=t_{i}(\tau).
\]
This test must succeed in order for the estimated HMM to be a valid
description of the metastable kinetics.

Note that a more general comparison is possible by comparing appropriate
norms of $\mathbf{T}^{\mathrm{pred}}(\tau)$ and $\mathbf{T}(\tau)$,
or, alternatively, of the correlation matrices $\mathbf{C}^{\mathrm{pred}}(\tau)=\boldsymbol{\chi}\widetilde{\boldsymbol{\Pi}}[\widetilde{\mathbf{T}}(\tau_{0})]^{n}\boldsymbol{\chi}^{\top}$
and $\mathbf{C}(\tau)=\boldsymbol{\Pi}\mathbf{T}(\tau)$. A practically
feasible comparison could be constructed in a similar way as the Chapman-Kolmogorow
test in \cite{PrinzEtAl_JCP10_MSM1}.

\section{Quantitative analysis}

One of the reasons for the great success of direct (non-hidden) Markov
models is that various quantities related to the molecule's thermodynamics,
kinetics and mechanisms can be calculated easily from an MSM transition
matrix. Therefore, although it will be shown in the applications section
that HMMs can be much superior to MSMs in modeling the kinetics, their
use needs to be motivated by showing that they are equally versatile
as MSM. This section goes through a number of commonly used molecular
observables and discusses how they can be computed from the HMM quantities
$\widetilde{\mathbf{T}}$ and $\boldsymbol{\chi}$. Interestingly,
for some observables, the computation from HMMs is even more straightforward
than from MSMs.

Some important quantities can be directly accessed through an eigenvalue
decomposition of the hidden transition matrix $\tilde{\mathbf{T}}(\tau)$.
Such a decomposition provides the eigenvalues $\tilde{\lambda}_{i}(\tau)$,
the right eigenvector matrix $\widetilde{\mathbf{R}}$, which contains
the right eigenvectors $\widetilde{\mathbf{r}}_{i}$ as columns, and
the left eigenvector matrix $\widetilde{\mathbf{L}}=\widetilde{\mathbf{R}}^{-1}$
which contains the left eigenvectors $\tilde{\mathbf{l}}_{i}^{T}$
as rows. The first left eigenvector can be normalized to a sum of
$1$, yielding the stationary distribution of hidden states, $\widetilde{\boldsymbol{\pi}}$. 

The stationary distribution $\widetilde{\boldsymbol{\pi}}$ provides
the probability of observing one of the metastable states. The free
energy of state $i$ with respect to an arbitrary reference state
$0$ is given by 
\[
\Delta F_{i}=-k_{B}T\ln\frac{\tilde{\pi}_{i}}{\tilde{\pi}_{0}}.
\]
Note that these free energy differences are associated to the weights
of the metastable states, even when the state space discretization
is poor. This is not the case when computing the free energy of metastable
states from an MSM, where a poor discretization can lead to significant
confusion which microstate should be associated to what degree to
a metastable state. However, when the stationary distribution is sought
on the microstates, it can be easily computed by transforming the
stationary distributions of metastable states through the output probabilities:
\[
\boldsymbol{\pi}=\boldsymbol{\chi}^{\top}\widetilde{\boldsymbol{\pi}}.
\]
A quantity of particular interest is the definition of the metastable
states themselves. In particular, which set of molecular structures
is metastable? This question has been an important driving force in
the development of MSMs. The original contribution in this field was
made by Schütte, Deuflhard and co-workers by noticing that for $m$
most metastable states, the signs of the dominant $m$ MSM eigenvectors
are indicative \cite{SchuetteFischerHuisingaDeuflhard_JCompPhys151_146,Deuflhard_LinAlgAppl_PCCA}.
Hence their PCCA method defined their metastable states as the set
of microstates with equal signs in the first MSM eigenvectors. A few
years later, Weber and Deuflhard have invented PCCA$+$ \cite{DeuflhardWeber_PCCA},
which is numerically and conceptually superior and assigns to each
microstate $i$ a membership $m_{ij}$ to each metastable state $j$
based on the proximity of microstate $i$ to a representative state
that is representative for metastable state $j$ in the space of the
dominant $m$ eigenvectors of the MSM transition matrix. While PCCA
and PCCA$+$ have nice theoretical properties, they are both unsatisfactory
from a statistical point of view. As the PCCA($+$) metastable states
are defined based on the transition matrix eigenvectors, any information
of the statistical significance is lost. Therefore, other methods
such as BACE \cite{Bowman_JCP12_BACE} have taken a Bayesian standpoint
and defined metastability based on information in the MSM transition
count matrix. HMMs directly provide information of the metastable
states. The output matrix $\boldsymbol{\chi}=[\chi_{ij}]$ directly
provides the probability that a given metastable state $i$ is observed
in a given microstate $j$. Its row vectors $\boldsymbol{\chi}_{i}$
therefore are probability distributions of metastable states on the
space of clusters. With the weight $\widetilde{\pi}_{i}$ these probability
distributions can be weighted, such that these vectors sum up to the
overall probability distribution of microstates: $\pi_{j}=\sum_{i}\widetilde{\pi}_{i}\chi_{ij}$.
Using Bayesian inversion, the $\boldsymbol{\chi}$ matrix can be transformed
into a membership matrix $\mathbf{M}=[m_{ij}]$ which contains the
information ``how much'' microstate $i$ belongs to metastable state
$j$:
\begin{eqnarray*}
\mathbf{M} & = & \boldsymbol{\Pi}^{-1}\boldsymbol{\chi}^{T}\widetilde{\boldsymbol{\Pi}}
\end{eqnarray*}
where $\boldsymbol{\Pi}=\mathrm{diag}(\boldsymbol{\pi})$ and $\widetilde{\boldsymbol{\Pi}}=\mathrm{diag}(\widetilde{\boldsymbol{\pi}})$.
This approach of defining metastable states unifies the advantages
of PCCA$+$ and of statistically-driven methods such as BACE: (1)
As in PCCA$+$, the memberships $\mathbf{M}$ are in the subspace
of the slow dynamics, and are therefore a mathematically meaningful
approach for characterizing metastability. (2) In contrast to PCCA$+$,
one does not need to define representative states. The membership
matrix $\mathbf{M}$ is a result of the HMM estimation itself. (3)
Since HMMs include $\boldsymbol{\chi}$ as a direct parameter, the
quantity $\boldsymbol{\chi}$ is directly amenable to statistical
treatment. When estimated via EM, $\boldsymbol{\chi}$ is the result
of a likelihood (or posterior) maximization. When using Monte-Carlo
sampling of the HMM likelihood \cite{ChoderaEtAl_BiophysJ11_BHMM},
the statistical uncertainty of elements in $\boldsymbol{\chi}$ can
be directly assessed.

Let us turn to kinetic quantities. The $m$ slowest relaxation rates,
or phenomenological rates, of the molecule are given by:
\begin{equation}
\tilde{\kappa}_{i}=-\frac{\ln|\tilde{\lambda}_{i}(\tau)|}{\tau}\label{eq_implied-rates}
\end{equation}
These rates, and their inverses, the relaxation timescales $\tilde{t}_{i}=\tilde{\kappa}_{i}$
are of central interest in kinetics, because they can be probed by
various experimental techniques. A main advantage of PMMs is that
- in stark contrast to MSMs - the rates $\tilde{\kappa}_{i}$ can
be estimated without systematic error. This is also true for HMMs,
when they are employed for metastable systems (see discussion in Sec.
\ref{sec_2}). From these relaxation rates, and the eigenvectors of
the hidden transition matrix, we can compute the rate matrix between
metastable states:
\begin{equation}
\widetilde{\mathbf{K}}=\widetilde{\boldsymbol{\Pi}}^{-1}\widetilde{\mathbf{L}}^{\top}\left[\begin{array}{cccc}
0\\
 & -\tilde{\kappa}_{2}\\
 &  & \ddots\\
 &  &  & -\tilde{\kappa}_{m}
\end{array}\right]\widetilde{\mathbf{L}}\label{eq_analysis_K}
\end{equation}
In contrast to MSMs, the transformation to a rate matrix is possible,
because the first $m$ processes are metastable, and therefore $\lambda_{1}...\lambda_{m}$
are positive such that the rates (\ref{eq_implied-rates}) exit.

Besides the decomposition into metastable states, and the rate or
transition matrix switching between them, the eigenvectors themselves
provide a quite direct understanding of the metastable dynamics: The
sign changes in $\widetilde{\mathbf{r}}_{i}$ and $\widetilde{\mathbf{l}}_{i}$
indicate structural changes that occur at the associated rates $\tilde{\kappa}_{i}$
or timescales $\tilde{t}_{i}$. On the discretized state space, these
eigenvectors occur as projections from the hidden states through the
output probability matrix: 
\begin{eqnarray}
\mathbf{s}_{i} & = & \mathbf{M}\widetilde{\mathbf{r}}_{i}\nonumber \\
\mathbf{q}_{i} & = & \boldsymbol{\chi}^{\top}\widetilde{\mathbf{l}}_{i}\label{eq_analysis_qi}
\end{eqnarray}
Note that these projected eigenvectors may significantly differ from
the right and left eigenvectors that are directly computed from an
MSM transition matrix on the cluster space. The projected eigenvectors
and the relaxation rates are the key components for calculating kinetic
experimental observables. In \cite{NoeEtAl_PNAS11_Fingerprints,KellerPrinzNoe_ChemPhysReview11},
we have derived general expressions for computing correlation and
relaxation experiments, that can be straightforwardly extended to
HMMs. In \cite{LindnerEtAl_JCP13_NeutronScatteringI} we have extended
this theory to scattering experiments.

An important source of kinetic information are time-correlation experiments.
These may be realized by taking trajectories from time-resolved single
molecule experiments, such as single molecule fluorescence or pulling
experiments, and computing time correlations from these trajectories.
Moreover, several ensemble kinetic experiments effectively measure
time-correlation functions, for example dynamical neutron scattering.
A general expression for modeling these experiments is that of the
time cross-correlation, of two experimentally observable quantities.
Given a partition into states $S_{i}$, let us denote by $a_{i}$
and $b_{i}$ the averages of the two experimentally observable quantities
over the discrete state $S_{i}$. $\mathbf{a}$, $\mathbf{b}$ are
the vectors with these averages as elements. The cross-correlation
for time $\tau$ can be expressed as:
\begin{eqnarray*}
\mathbb{E}[a(t)b(t+\tau)] & = & \sum_{i=1}^{m}\mathrm{e}^{-\tau\kappa_{i}}\langle\mathbf{a},\mathbf{q}_{i}\rangle\langle\mathbf{b},\mathbf{q}_{i}\rangle\\
 & = & \langle\mathbf{a},\boldsymbol{\pi}\rangle\langle\mathbf{b},\boldsymbol{\pi}\rangle+\sum_{i=2}^{m}\mathrm{e}^{-\tau\kappa_{i}}\langle\mathbf{a},\mathbf{q}_{i}\rangle\langle\mathbf{b},\mathbf{q}_{i}\rangle
\end{eqnarray*}
Autocorrelation experiments can be modeled by simply setting $a=b$. 

Alternatively, relaxation experiments can be used to probe the molecules'
kinetics. In these experiments, the system is allowed to relax from
a nonequilibrium starting state with probability distribution. Examples
are temperature-jump, pressure-jump, or pH-jump experiments, rapid
mixing experiments, or experiments where measurement at $t=0$ starts
from a synchronized starting state, such as in processes that are
started by an external trigger like a photoflash. We consider initial
distributions that are modeled on the metastable states, $\widetilde{\mathbf{p}}(0)$.
For example, in an ideal two-state folder, the relaxation experiment
shifts probabilities between the two metastable states, and a meaningful
value of $\widetilde{\mathbf{p}}(0)$ could be computed from available
experimental titration curves. The time evolution of such an initial
distribution can be computed by propagating it with the transition
or rate matrix that describe the dynamics for the conditions after
the trigger: 
\begin{eqnarray*}
\mathbf{p}_{\tau}^{\top} & = & \tilde{\mathbf{p}}_{0}^{\top}[\widetilde{\mathbf{T}}(\tau_{0})]^{n}\boldsymbol{\chi}\\
 & = & \tilde{\mathbf{p}}_{0}^{\top}\exp[\tau\widetilde{\mathbf{K}}]\boldsymbol{\chi}
\end{eqnarray*}
with $\tau=n\tau_{0}$. The ensemble average $\mathbb{E}_{\boldsymbol{p}(0)}[a(\tau)]$
of an experimentally measurable quantity, $a$, is recorded while
the system relaxes from the initial distribution $\widetilde{\mathbf{p}}(0)$
to the new equilbrium distribution $\widetilde{\boldsymbol{\pi}}$.
The expectation value of the signal at time $\tau$ is then given
by
\[
\mathbb{E}_{\widetilde{\mathbf{p}}_{0}}[a(\tau)]=\sum_{i=1}^{m}\mathrm{e}^{-\tau\kappa_{i}}\langle\mathbf{a},\mathbf{q}_{i}\rangle\langle\tilde{\mathbf{l}}_{i},\widetilde{\mathbf{p}}_{0}^{*}\rangle
\]

where $\widetilde{\mathbf{p}}{}_{0}^{*}$ is the \emph{excess probability
distribution $\widetilde{\mathbf{p}}{}_{0}^{*}=\Pi^{-1}\widetilde{\mathbf{p}}{}_{0}$.
$\mathbb{E}_{\widetilde{\mathbf{p}}_{0}}[a(\tau)]$} is again a multiexponential
decay function with amplitudes $\langle\mathbf{a},\mathbf{q}_{i}^{l}\rangle\langle\tilde{\mathbf{l}}_{i},\widetilde{\mathbf{p}}_{0}^{*}\rangle$.
Each of the amplitudes is associated with an eigenvector of the transition
matrix, and therefore readily interpretable in terms of structural
changes.

The combination of Markov models and the spectral theory given is
useful to compare simulations and experiments via the \emph{dynamical
fingerprint} representation of the system kinetics \cite{NoeEtAl_PNAS11_Fingerprints}.
Furthermore, this approach permits to design experiments that are
optimal to probe individual relaxations \cite{NoeEtAl_PNAS11_Fingerprints}.

Finally, detailed molecular mechanisms of a particular process that
transitions between two states $A$ and $B$ can be calculated with
transition path theory (TPT) \cite{EVandenEijnden_TPT_JStatPhys06}.
Here, $A$ and $B$ may be associated to the unfolded and folded ensembles
in protein folding, or to the dissociated and assocatied states in
protein-ligand binding. TPT can be straightforwardly applied on the
level of metastable states. This is done either by directly applying
to the transition matrix $\widetilde{\mathbf{T}}$ (see \cite{NoeSchuetteReichWeikl_PNAS09_TPT}),
or by computing rate matrix $\tilde{\mathbf{K}}$ (see above) and
conducting TPT as described in \cite{MetznerSchuetteVandenEijnden_JCP06_TPT}.

\section{Applications}

\begin{figure}
\includegraphics[width=0.5\columnwidth]{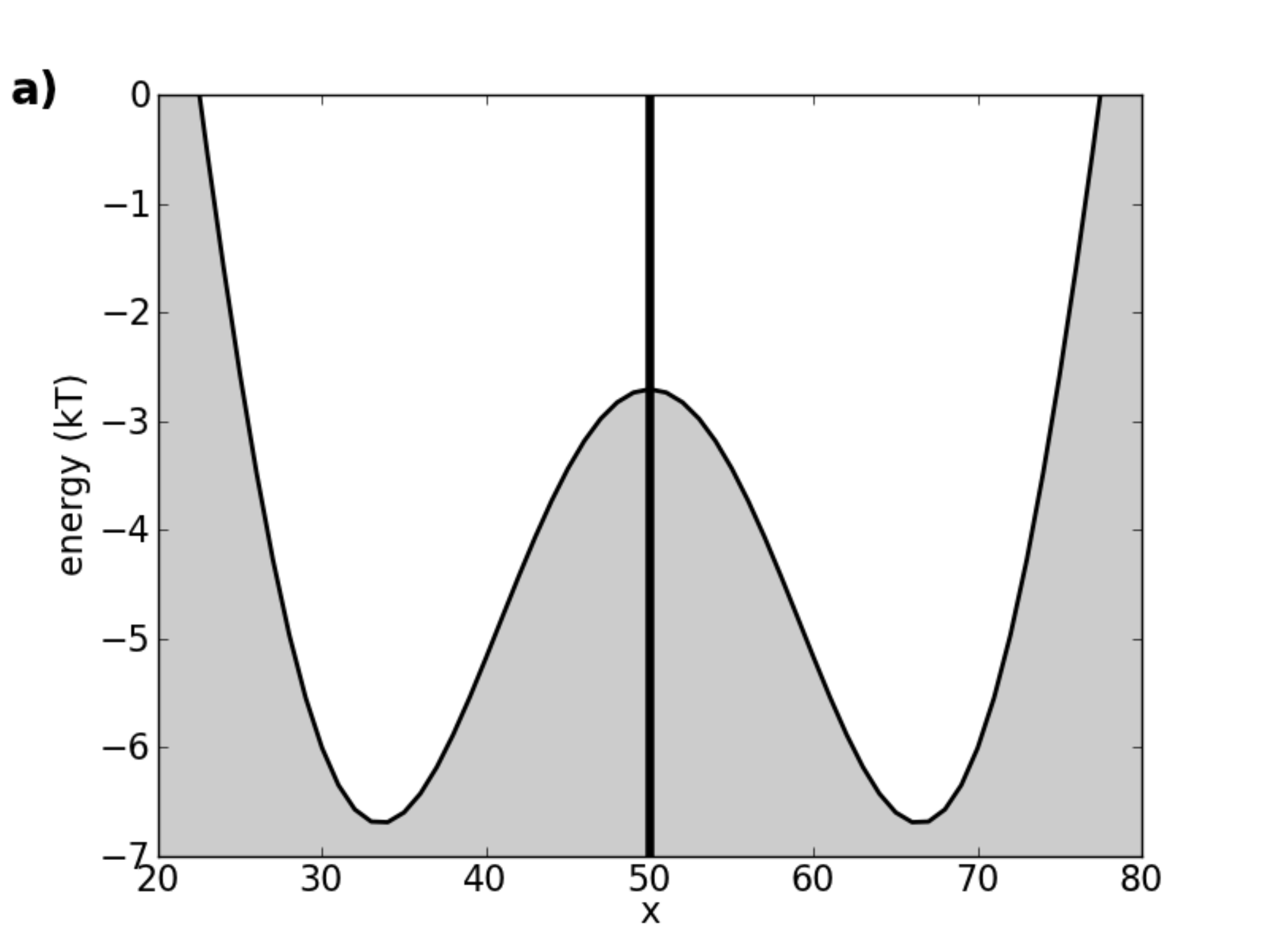}\includegraphics[width=0.5\columnwidth]{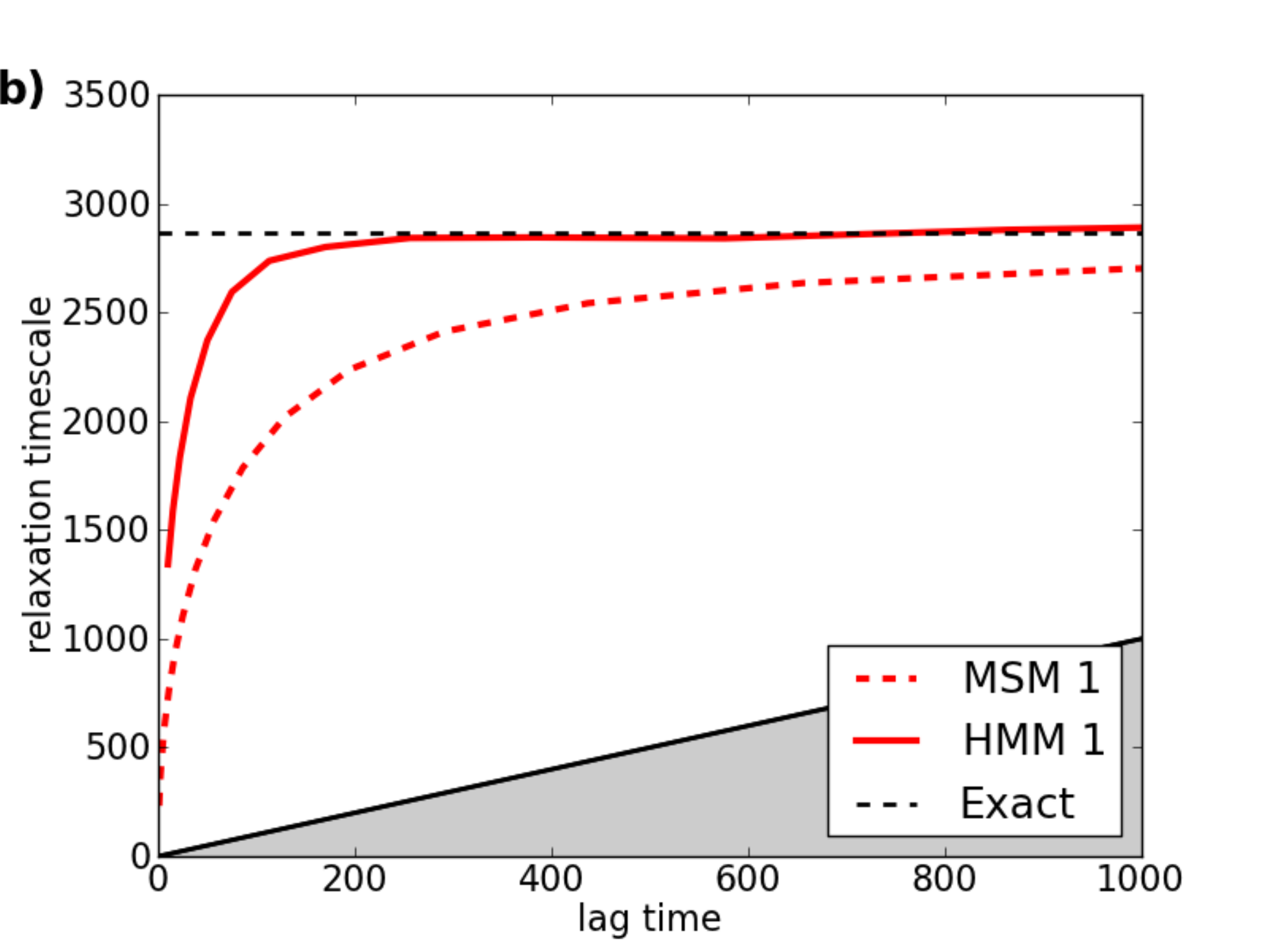}

\includegraphics[width=0.5\columnwidth]{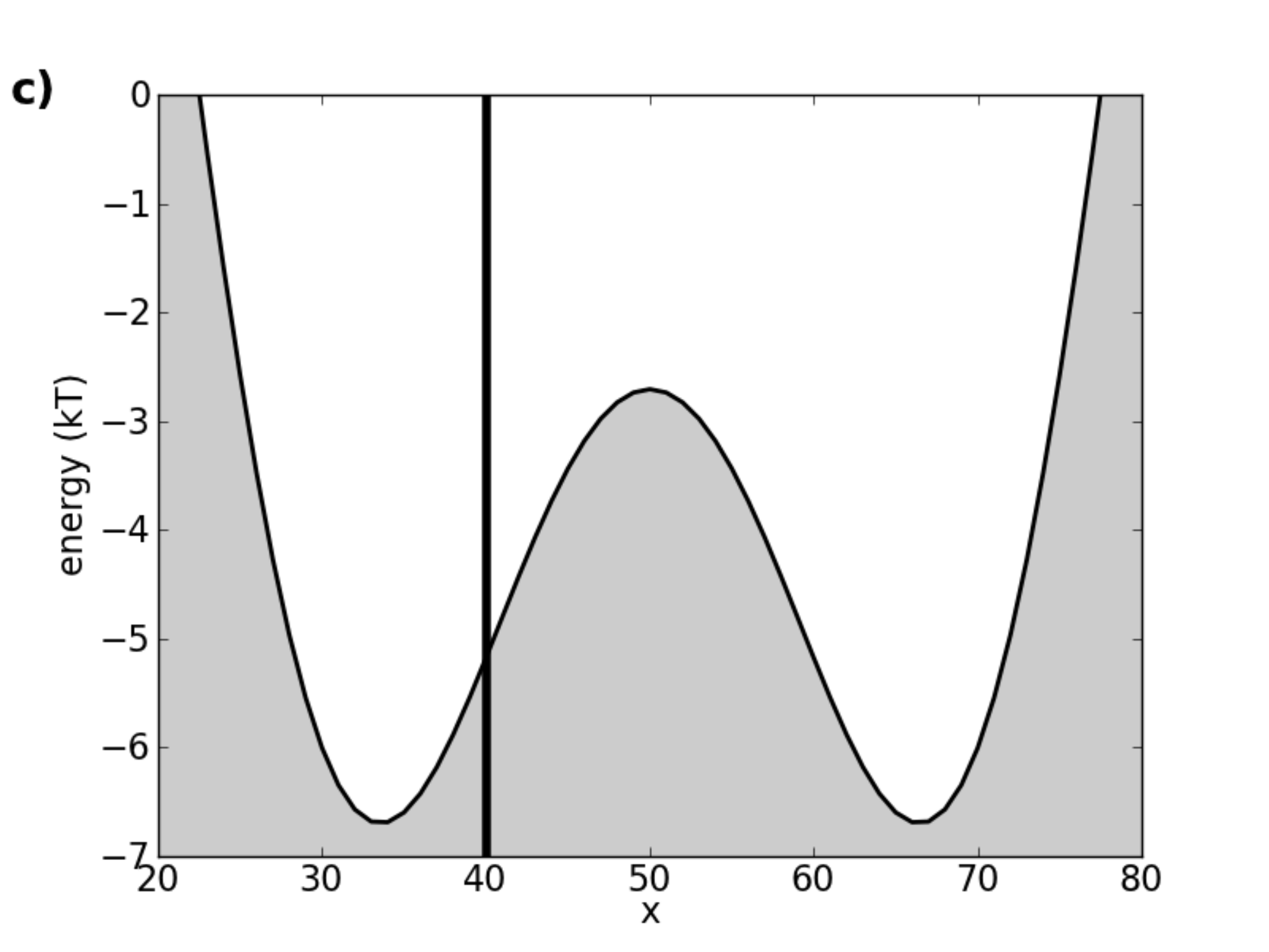}\includegraphics[width=0.5\columnwidth]{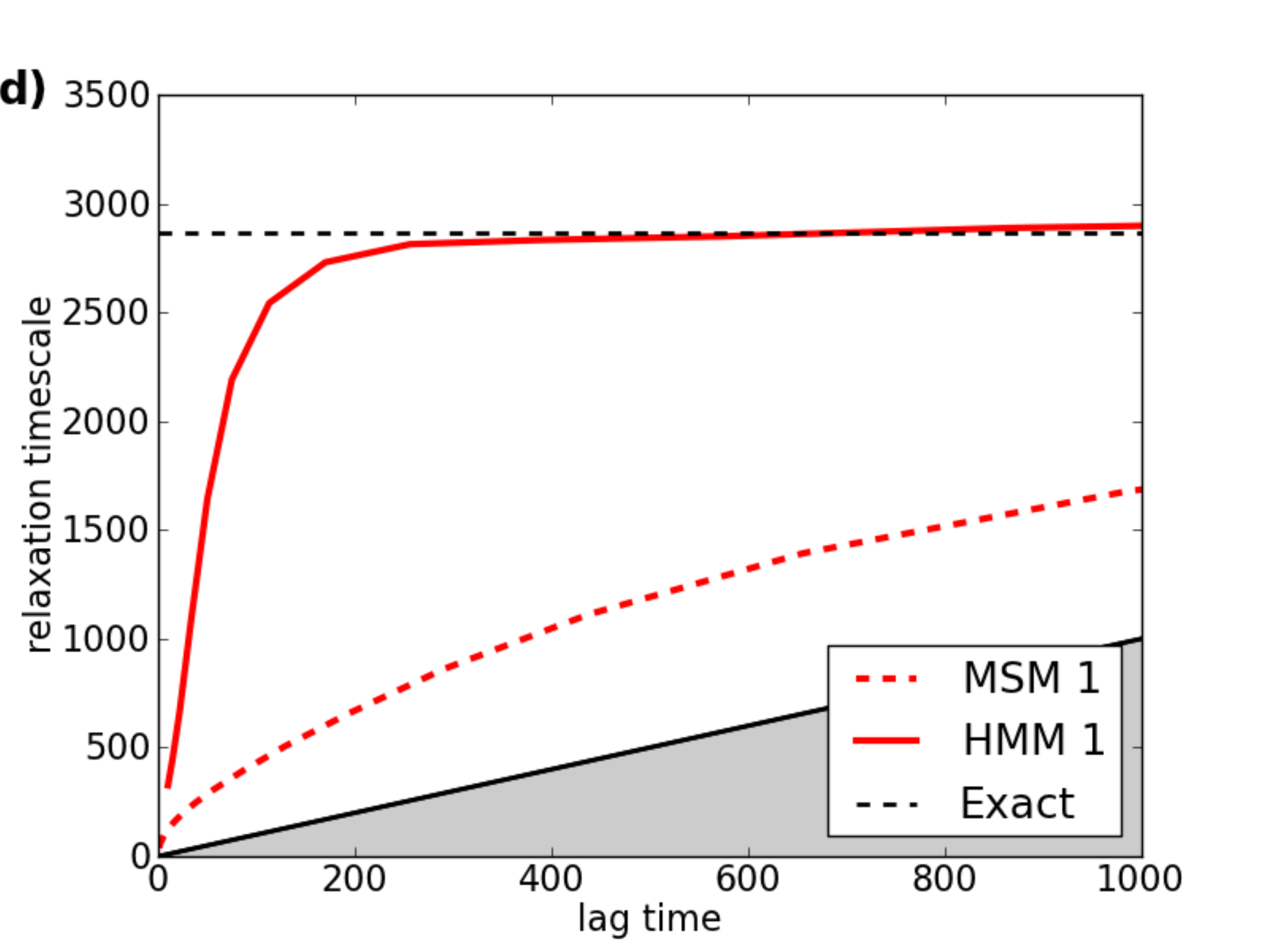}

\caption{\label{fig1_2state}Comparison of MSM and PMM/HMM for modeling the
diffusion in a bistable potential, using (a) good and (c) poor discretization
into two states. b,c): $\tau$-dependence of relaxation timescales
computed with MSMs and HMMs. The grey region is the $\tau>t$ region
where no numerically robust estimation of the relaxation timescale
$t$ is possible.}

\end{figure}

Fig. \ref{fig1_2state} compares the performances of MSMs and HMMs
on a diffusion in a metastable two-well potential (model details see
SI of \cite{PrinzEtAl_JCP10_MSM1}). The spate space is discretized
into two clusters, comparing results for a good discretization separating
the two metastable states in the transition region (Fig. \ref{fig1_2state}a)
and a bad discretization that splits one of the metastable states
(Fig. \ref{fig1_2state}c). For the good discretization, the MSM estimate
converges to the true timescale when the lag time $\tau$ is increased,
although slowly with an error that vanishes as $\tau^{-1}$ --- see
\cite{PrinzChoderaNoe_PRL11_RateTheory} for derivation. For the poor
discretization, the convergence of the MSM is so slow that it does
not come close to the true timescale before hitting the ``forbidden''
region $\tau>t$ at which no numerically robust MSM estimate is possible
\cite{BeauchampEtAl_PNAS2012_FoldingMSMs}. In contrast, the PMM/HMM
converges quickly to the true timescale, and the timescale estimate
then stays constant in $\tau$. The speed of this convergence goes
exponential with the greatest neglected timescale, as $\exp(-\tau/t_{3})$.
Thus, the HMM behaves as a multi-$\tau$ estimator analyzed in \cite{PrinzChoderaNoe_PRL11_RateTheory}.
Obtaining a good model for the slow kinetics for a short lag time
$\tau$ is very important for ensemble simulations, because it allows
to keep the length of the individual simulations short as well. Shorter
trajectory lengths also permit a more rapid turnover in adaptive sampling
simulations \cite{SinghalPande_JCP123_204909,Singhal_JCP07,BowmanEnsignPande_JCTC2010_AdaptiveSampling},
thus allowing to get statistically converged estimates of the slow
kinetics with lesser total sampling effort.

\begin{figure}
\includegraphics[width=1\columnwidth]{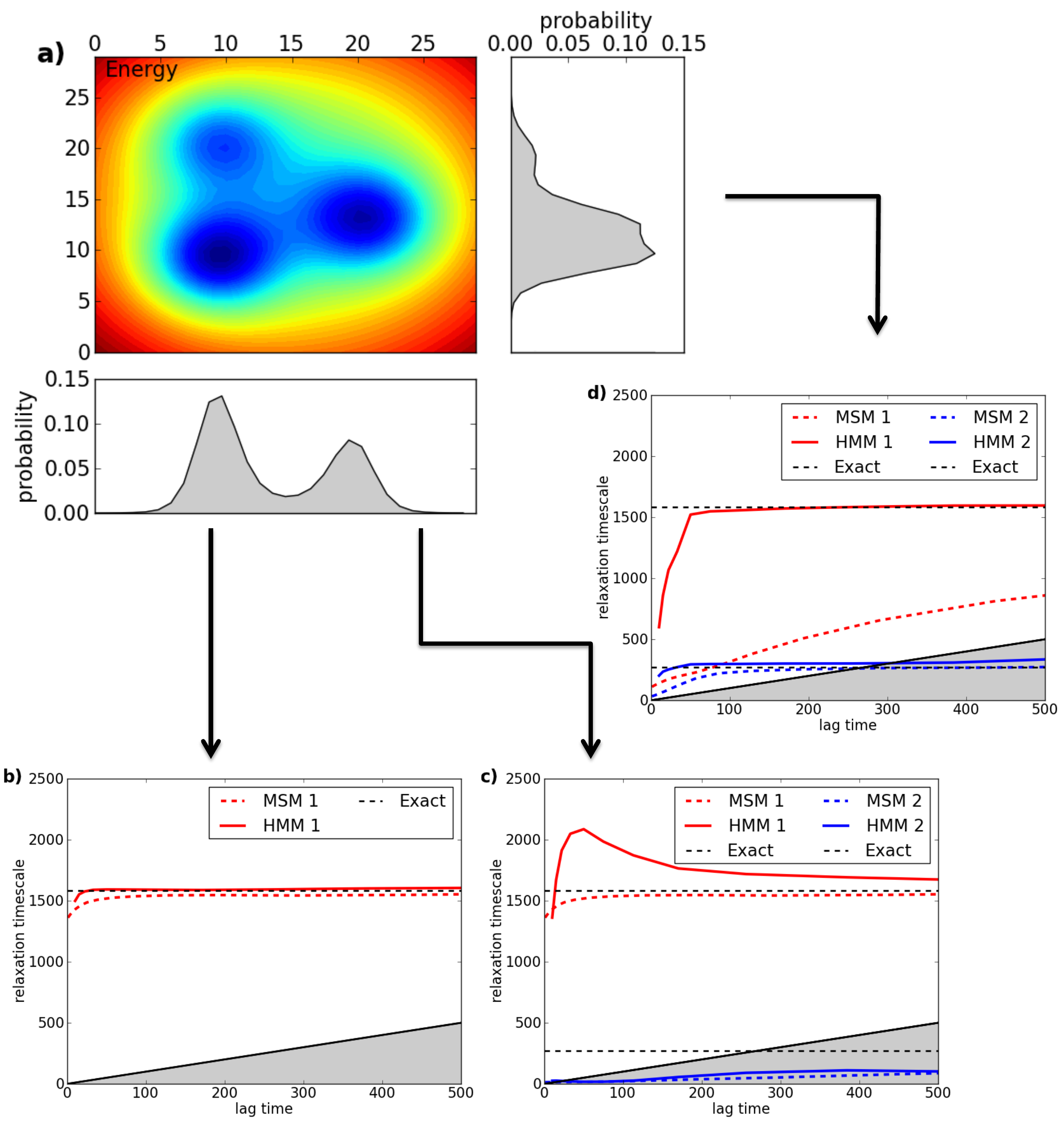}

\caption{\label{fig2_3state}Comparison of MSM and PMM/HMM for modeling the
diffusion in a bistable potential (model details see SI of \cite{PrinzEtAl_JCP10_MSM1})
from projections onto the $x$- and $y$-coordinate, respectively.
A fine (30-state) discretization in the respectively observed coordinate
is used in order to build the MSM or PMM/HMM. a) Energy landscape
and observed probability densities in $x$ and $y$. b,c) 1- and 2-timescale
estimate for the projection onto $x$. d) Timescale estimation for
the projection onto $y$.}
\end{figure}

Fig. \ref{fig2_3state} compares the performances of MSMs and HMMs
when constructing the model on a subspace of conformation space that
neglects important degrees of freedom. For the diffusive dynamics
in the two-dimensional three-well potential shown in Fig. \ref{fig2_3state}a,
both dimensions are needed in order to separate the three metastable
states from another. The projections of the probability density onto
either the $x$ or $y$ coordinate (grey distributions) only exhibit
two poorly separated modes. The slowest relaxation timescale is associated
to the transition between the two deep wells, and therefore mostly
with the $x$-axis, while the second-slowest relaxation timescale
is associated to the transition between the lower left minimum and
the shallow upper minimum. An MSM is able to estimate the slowest
relaxation timescale from the $x$-projections (Fig. \ref{fig2_3state}b,c),
and the second-slowest relaxation timescale from the $y$-projection
(Fig. \ref{fig2_3state}d). However, an MSM is not able to estimate
both slow processes simultaneously. The HMM performs similarly on
the $x$-projection: When using a two-state HMM, the slowest timescale
is estimated very accurately, and with a shorter lagtime than the
MSM (Fig. \ref{fig2_3state}a). When using a three-state HMM, the
result for the slowest timescale actually gets worse (Fig. \ref{fig2_3state}b),
while still being unable to estimate the second-slowest timescale.
This shows a limitation of the method in the worst-case scenario that
a process is completely hidden: For the $x$-projection, the stable
lower left state and the less stable upper state are projected to
exactly the same observable values. Since the upper state only exchanges
with the lower state and has much shorter lifetimes, its presence
does not even affect the kinetics significantly. Therefore, the projection
onto $x$ really behaves like a system with two-state kinetics, and
using in a three-state HMM will compromise the estimation. 

How can the estimate become worse when \emph{too many} hidden states
are used? The answer lies in the structure the hidden HMM transition
matrix which has eigenvectors $\tilde{\mathbf{l}}_{i}$ associated
with the slowest processes. When analyzing a two-state system with
two hidden states, the HMM transition matrix will have be two two-element
vectors $\tilde{\mathbf{l}}_{1}=\widetilde{\boldsymbol{\pi}}$ and
$\tilde{\mathbf{l}}_{2}$, associated with the stationary distribution
and the slowest relaxation process, respectively, and these eigenvectors
will fulfill the orthogonality condition $\langle\tilde{\mathbf{l}}_{2},\widetilde{\boldsymbol{\pi}}\rangle=0$.
When analyzing a two-state system with three eigenvectors, the transition
matrix will have a third eigenvector $\tilde{\mathbf{l}}_{3}$, but
there is no relaxation process in the data associated to that. Therefore,
the HMM estimate will produce a random vector for $\tilde{\mathbf{l}}_{3}$.
Unfortunately, this also affects the quality of the other eigenvectors
$\tilde{\mathbf{l}}_{1}=\widetilde{\boldsymbol{\pi}}$ and $\tilde{\mathbf{l}}_{2}$,
because these eigenvectors are linked by pairwise orthogonality constraints.
The MSM is less affected by this problem, because it has many more
($n$) eigenvectors, so errors in estimating the fast process eigenvectors
do not necessarily compromise the slow process eigenvectors. This
emphasizes that it is important to use HMMs in the right setting:
estimating an HMM with $m$ states requires $m$ relaxation processes
to be present in the data, \emph{and} having a timescale separation
to the $(m+1)$th process. 

Fig. \ref{fig2_3state}c shows that the three-state HMM is able to
accurately estimate both relaxation timescales from the $y$-projection,
and is therefore superior to the MSM in this case.

\begin{figure*}[!t]
\begin{minipage}[b][1\totalheight][t]{0.4\textwidth}%
\includegraphics[width=1\columnwidth]{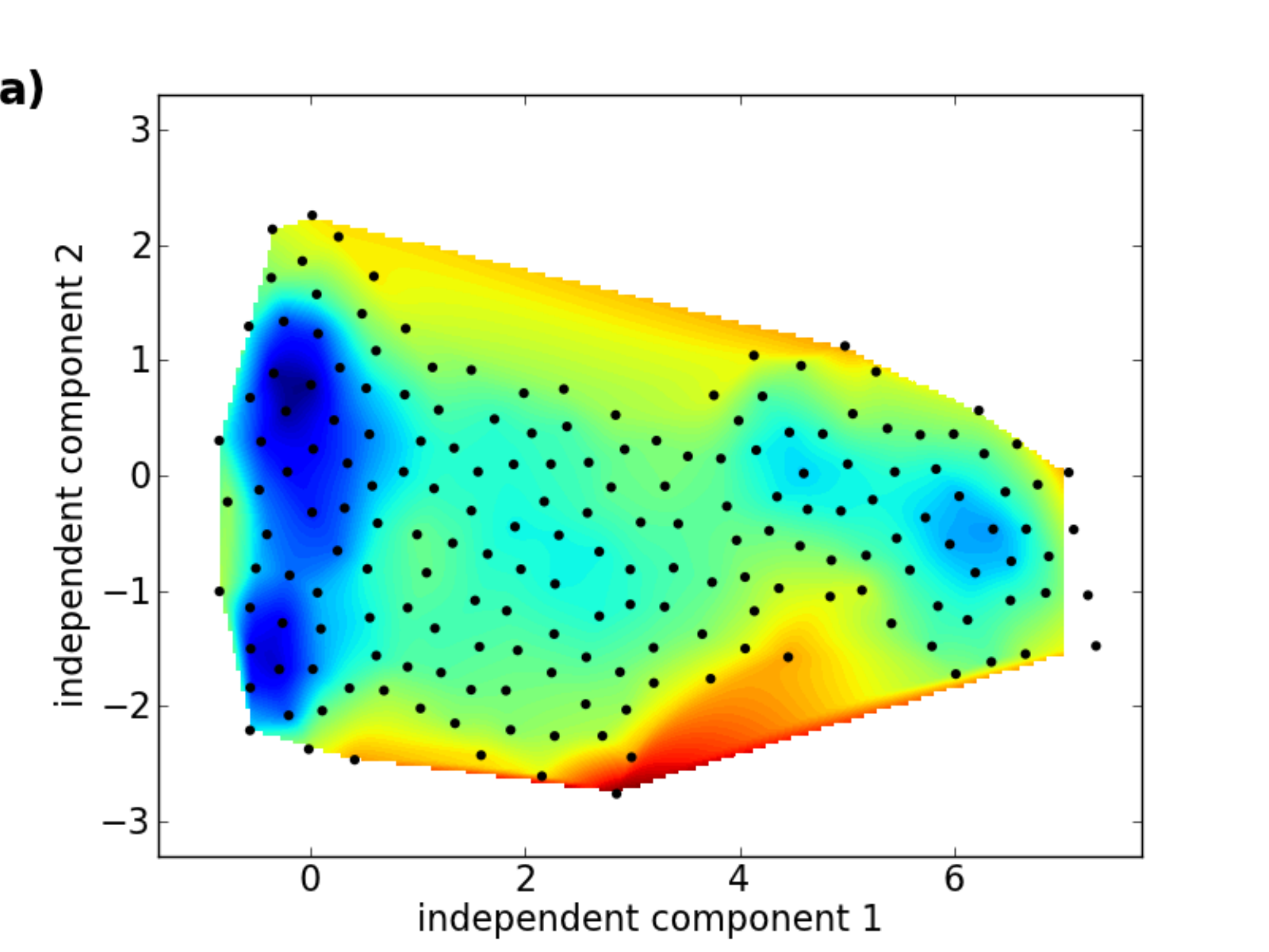}

\includegraphics[width=0.5\columnwidth]{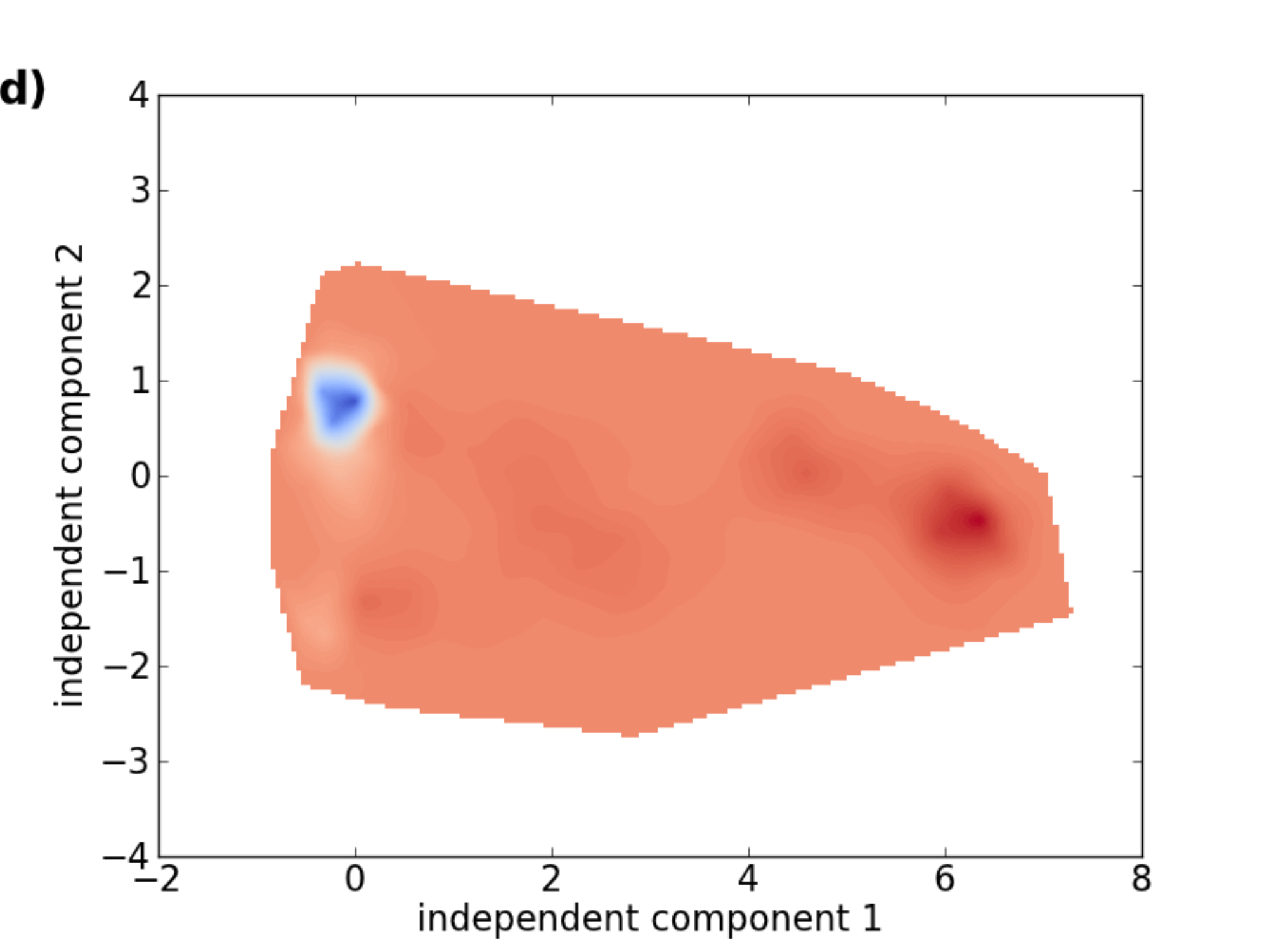}\includegraphics[width=0.5\columnwidth]{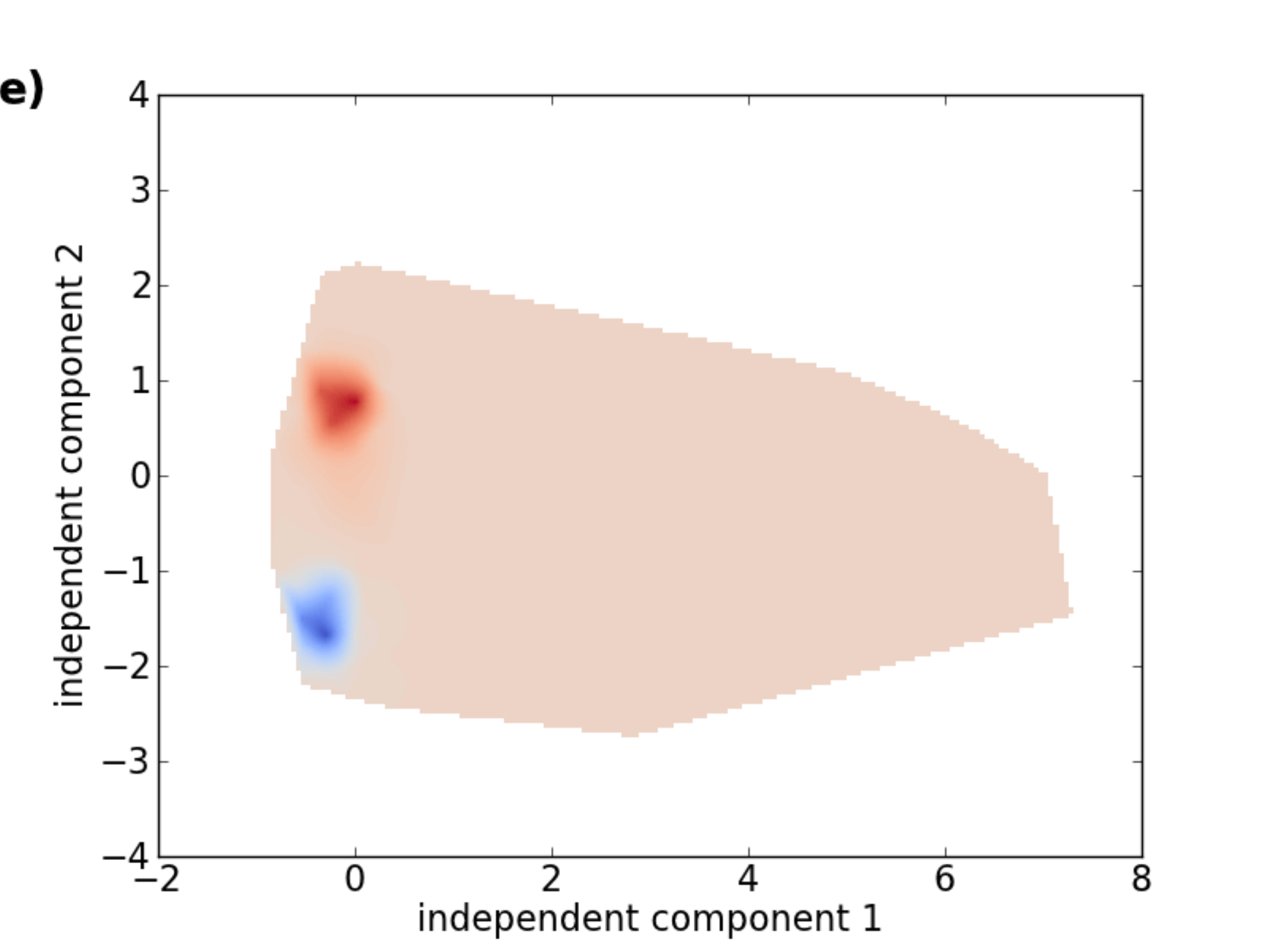}%
\end{minipage}%
\begin{minipage}[b][1\totalheight][t]{0.6\textwidth}%
\includegraphics[width=0.5\columnwidth]{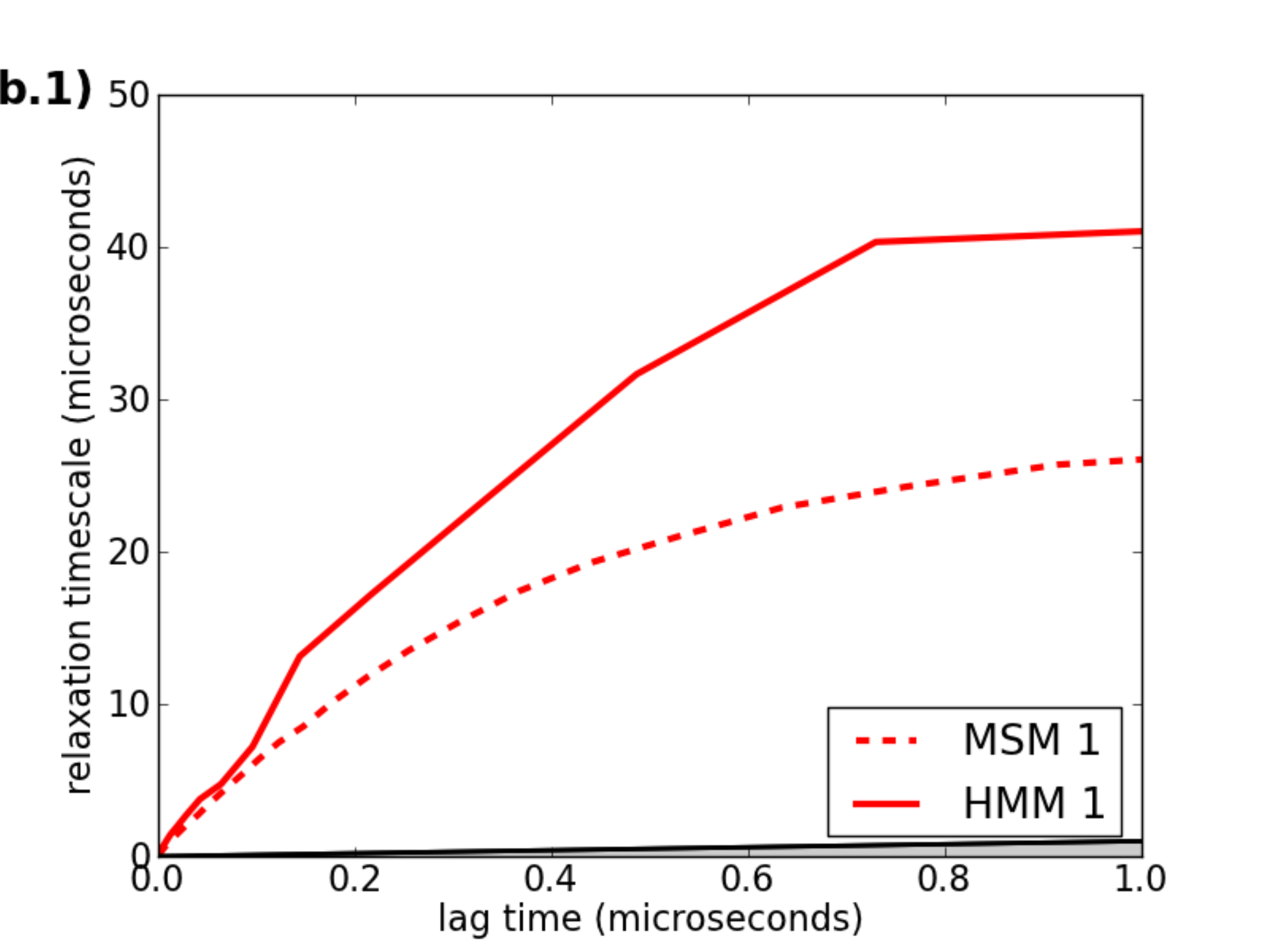}\includegraphics[width=0.5\columnwidth]{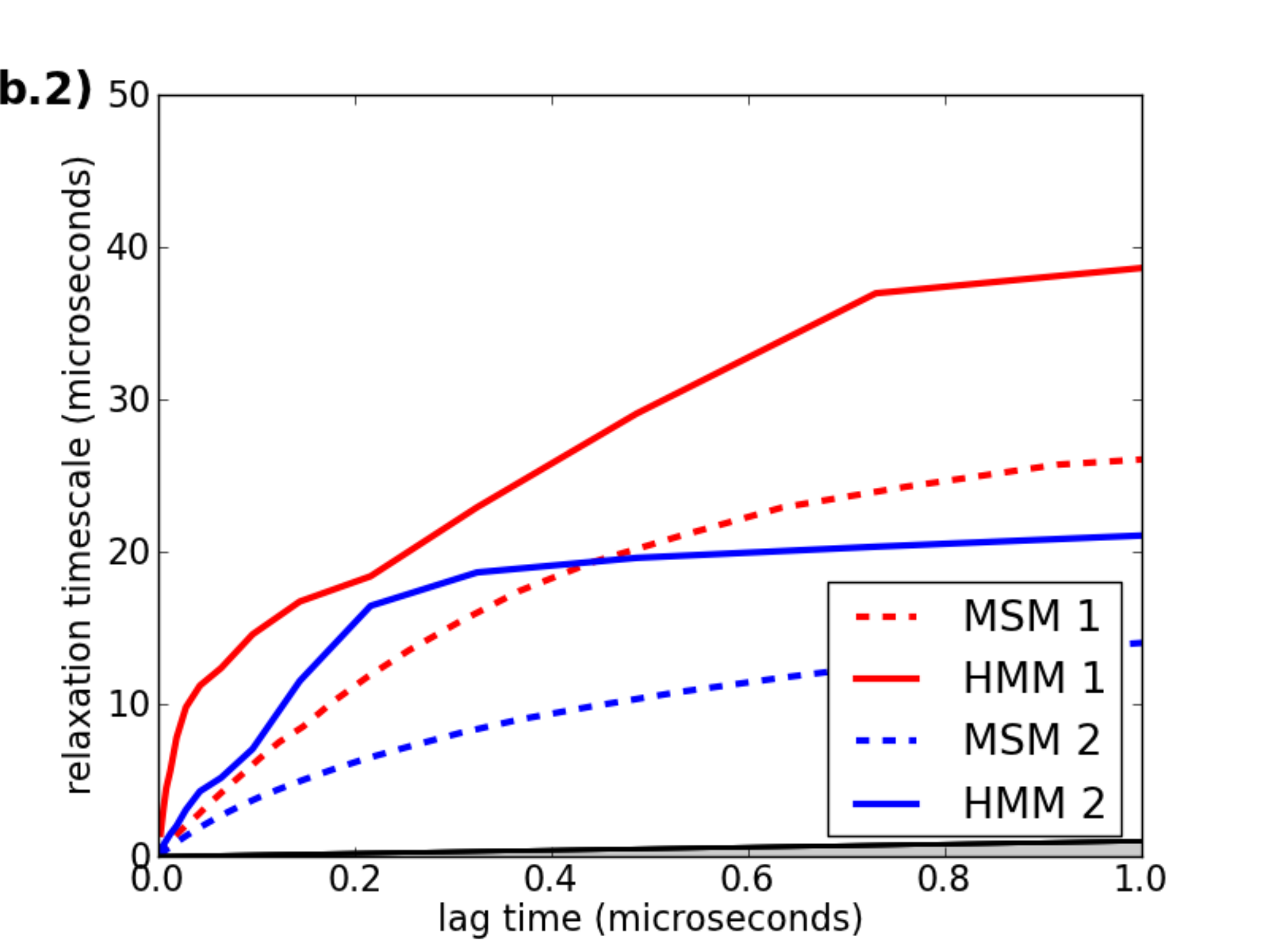}
\includegraphics[width=0.5\columnwidth]{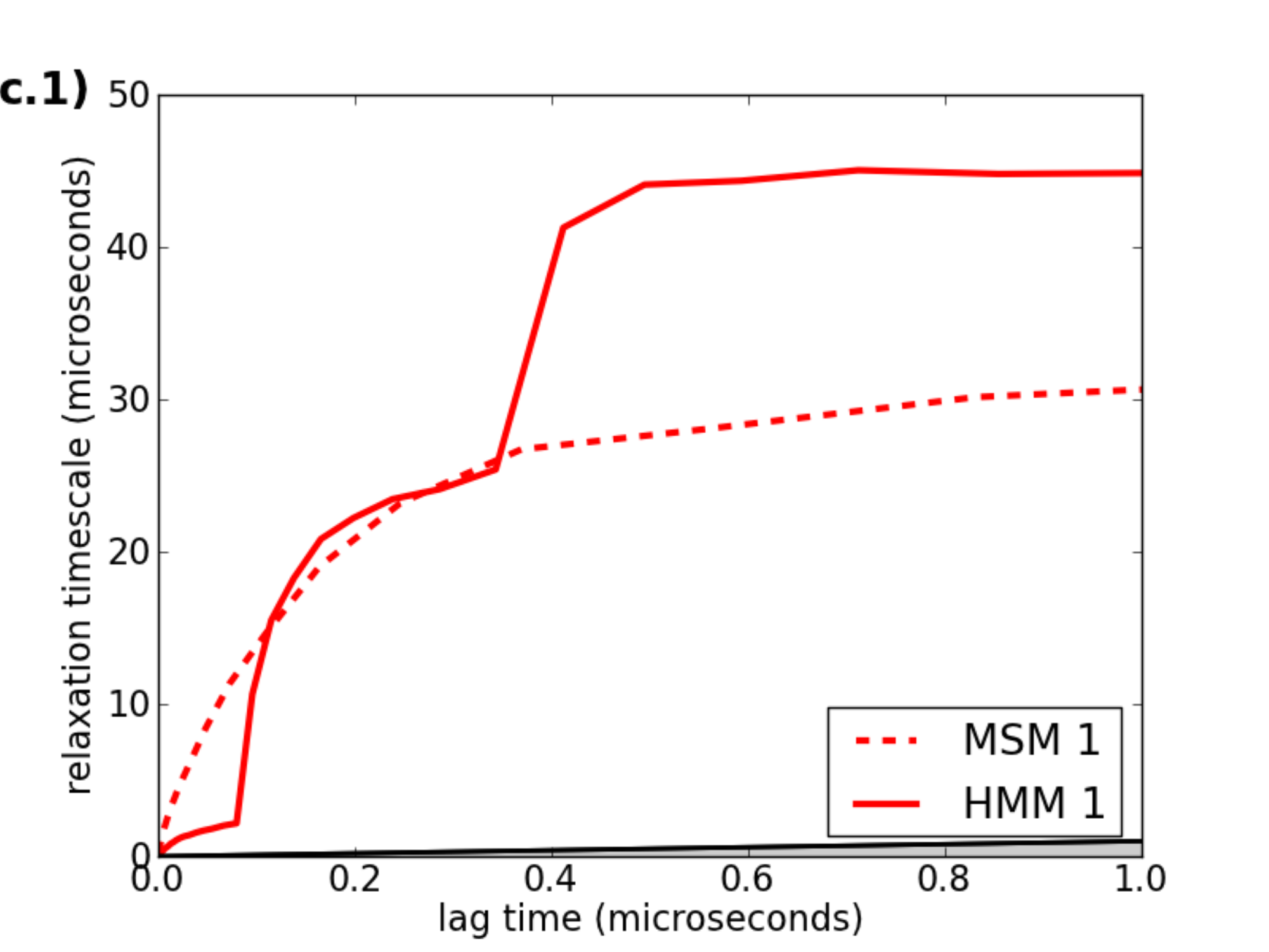}\includegraphics[width=0.5\columnwidth]{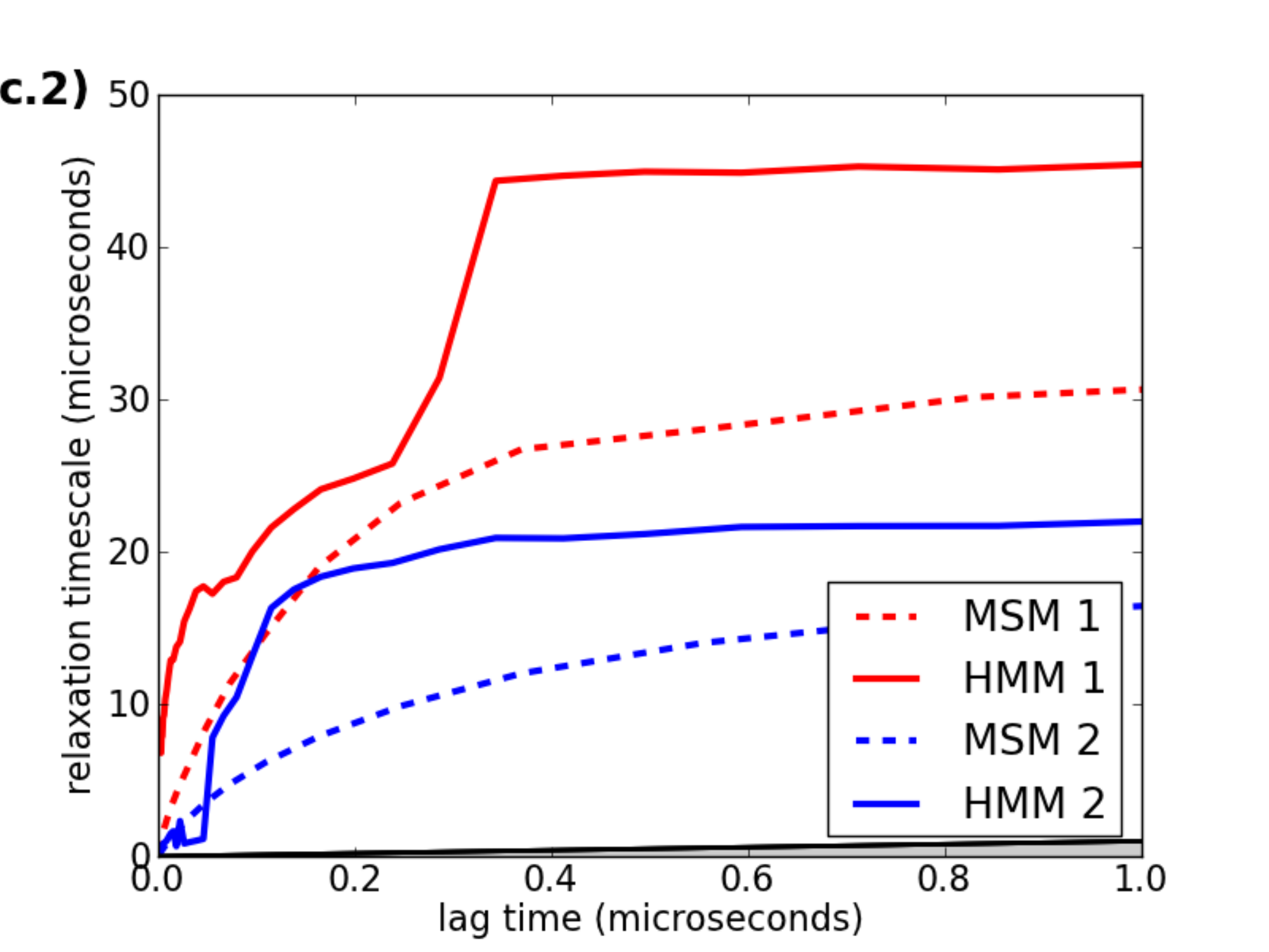}%
\end{minipage}

\includegraphics[width=1\textwidth]{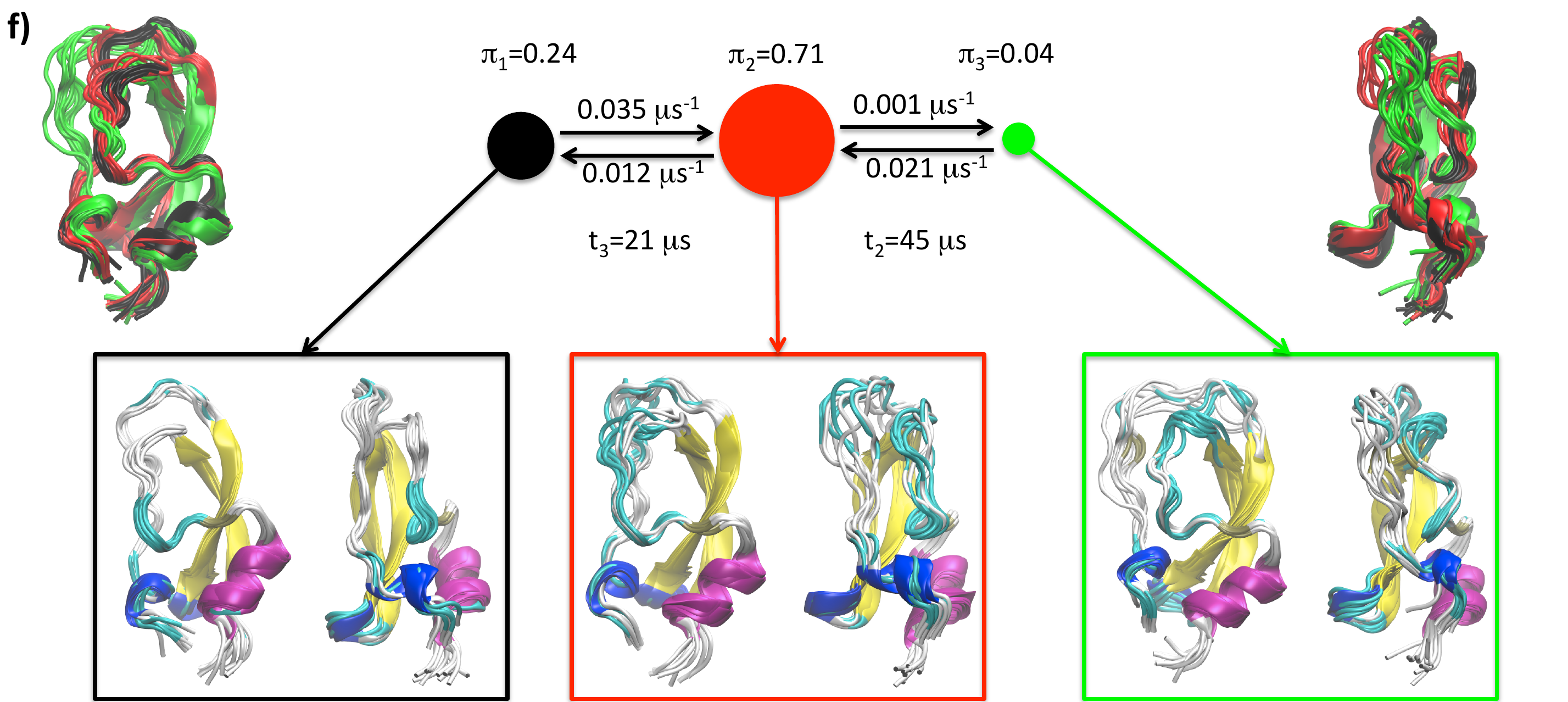}

\caption{\label{fig3_BPTI}Comparison of MSM and PMM/HMM for modeling the conformation
dynamics of BPTI using a the 1 ms simulation trajectory generated
by the Anton supercomputer \cite{Shaw_Science10_Anton} and kindly
provided by D.E. Shaw research. The discretization consisted of either
13 or 191 clusters, approximately uniformly distributed on the data
projected onto the space of the two dominant independent components
of a TICA analysis (see \cite{PerezEtAl_JCP13_TICA} for details),
using the EMMA implementation \cite{SenneSchuetteNoe_JCTC12_EMMA1.2}.
a) 191 clusters and a visualization of their free energy $-\ln\pi_{i}$.
b,c) Comparison of the slowest MSM timescales with the timescales
of the 2 and 3-state HMM, respectively, using 13 clusters (b.1,c.1)
or 191 clusters (b.2,c.2) . d,e) Visualization of the second and third
eigenvectors, $\mathbf{l}_{2}$ and $\mathbf{l}_{3}$ for the 191
cluster discretization. f) Three-state rate matrix corresponding to
the 3-state HMM. The structures are overlays of 10 frames drawn from
the state distributions $\boldsymbol{\chi}_{1}$, $\boldsymbol{\chi}_{2}$,
and $\boldsymbol{\chi}_{3}$.}
\end{figure*}

Fig. \ref{fig3_BPTI} shows the analysis of a 1 ms MD simulation of
the bovine pancreatic trypsin inhibitor (BPTI) produced on the Anton
supercomputer \cite{Shaw_Science10_Anton} and kindly provided by
D.E. Shaw research. We again consider two and three hidden states,
because an MSM analysis suggested gaps after the slowest and second-slowest
timescale. To obtain a cluster discretization, we first computed the
slowest independent components with time-lagged independent component
analysis (TICA) \cite{Molgedey::94} as described in \cite{PerezEtAl_JCP13_TICA}
using the EMMA1.4 implementation. The data was then projected onto
the two slowest components and we considered two cluster discretizations,
into 13 and 191 clusters. Fig. \ref{fig3_BPTI} shows a scatter plot
of the 191 cluster centers in the two dominant independent components.
The color code is a map of the logarithmized probability map of the
clusters, indicating a free energy surface. Note that such free-energy
surfaces generally suffer from overlap of states in the directions
not resolved in this plot, and only serves to provide a qualitative
impression where the regions with most statistical weight are. Fig.
\ref{fig3_BPTI}b,c show that the MSMs slowly converge towards slowest
timescale estimates of around 30 and 15 $\mu s$, while the HMMs converge
to robust and nearly $\tau$-constant estimates of timescales around
40 and 20 $\mu s$ - at lagtimes of 0.7 $\mu s$ for the 13-cluster
partition and at a lagtime of 0.3 $\mu s$ for the 191-cluster partition.
The HMMs therefore estimate somewhat larger relaxation timescales
and do that robustly for shorter lag times. Fig. \ref{fig3_BPTI}c.1
nicely shows what happens when employing a two-state HMM in a three-state
kinetics system: for short lagtimes, the HMM first finds the faster
timescale, and after a lagtime of about $\tau=0.3\:\mu s$ then jumps
to the slower timescale.

Fig. \ref{fig3_BPTI}d,e illustrate the two slow processes by plotting
the projected eigenvectors $\mathbf{q}_{1}$ and $\mathbf{q}_{2}$
(Eq. \ref{eq_analysis_qi}) on the two dominant independent components.
The slowest process is associated with probability exchange along
the first independent component (\ref{fig3_BPTI}d) and the second-slowest
process along the second independent component (\ref{fig3_BPTI}e).
\ref{fig3_BPTI}f illustrates the structures associated with the three
corresponding metastable states by plotting overlays of 10 structures
each selected from the metastable state output distributions $\{\boldsymbol{\chi}_{1},\boldsymbol{\chi}_{2},\boldsymbol{\chi}_{3}\}$
that are directly estimated by the HMM. Here, the black state is associated
with the lower left minimum in Fig. \ref{fig3_BPTI}a, and is the
most ordered structure. The red state is associated with the top left
minimum in Fig. \ref{fig3_BPTI}a, and is a slightly less ordered
structure, while the green state is associated with the rightmost
minimum in Fig. \ref{fig3_BPTI}a and exhibits a re-folded loop on
the $N$-terminal side of the backbone. 

Fig. \ref{fig3_BPTI}f also shows the $3\times3$ rate matrix between
metastable states computed from Eq. (\ref{fig3_BPTI}f). This shows
that the three metastable states are linearly connected, with the
black and red states exchanging on the faster 20 $\mu s$ timescale,
while the red state and the green state exchange on the slower 40
$\mu s$ timescale. Note that the green state is rather unstable,
and actually only one transition into and back out of the green state
occurs in the 1 ms trajectory, while the red and black states interchange
more frequently. Therefore the 40 $\mu s$ timescale is dominated
by the relatively short exit time from the green state, and this process
is statistically unreliable - thus the 40 $\mu s$ timescale is a
rough estimate. It is possible to extend the present HMM estimations
towards a fully Bayesian approach (analogously to \cite{ChoderaEtAl_BiophysJ11_BHMM}).
Thus, in the future, it will be possible to compute error bars on
the HMM estimates.

\begin{figure}[t]
a)\includegraphics[width=0.2\columnwidth]{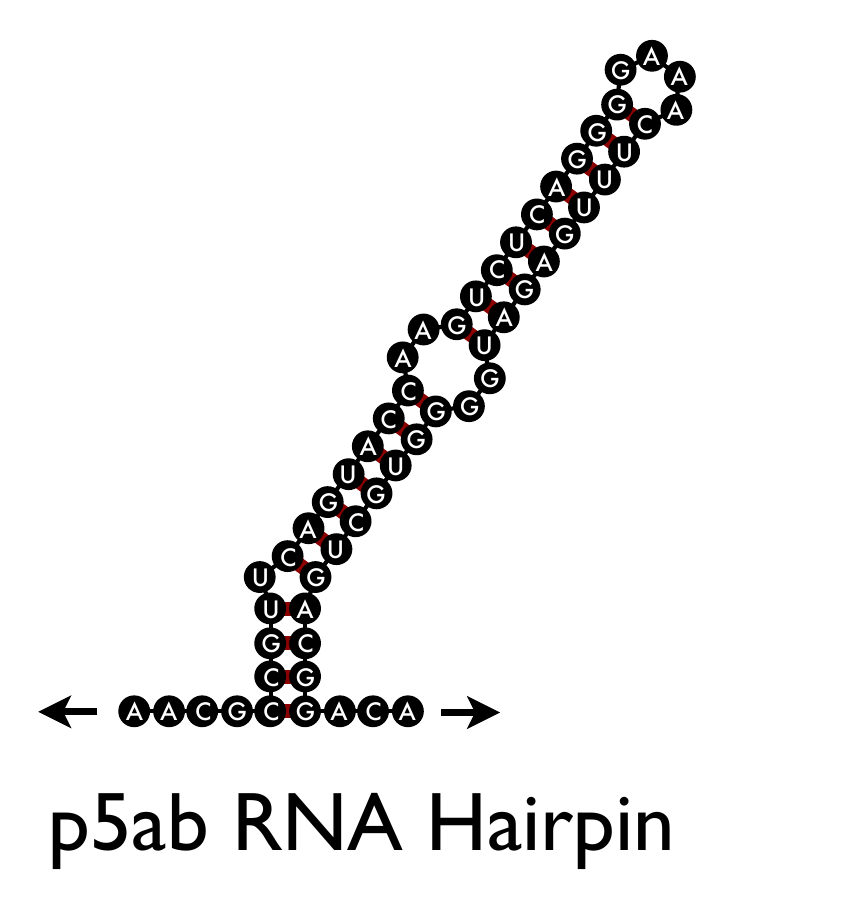} b)\includegraphics[width=0.7\columnwidth]{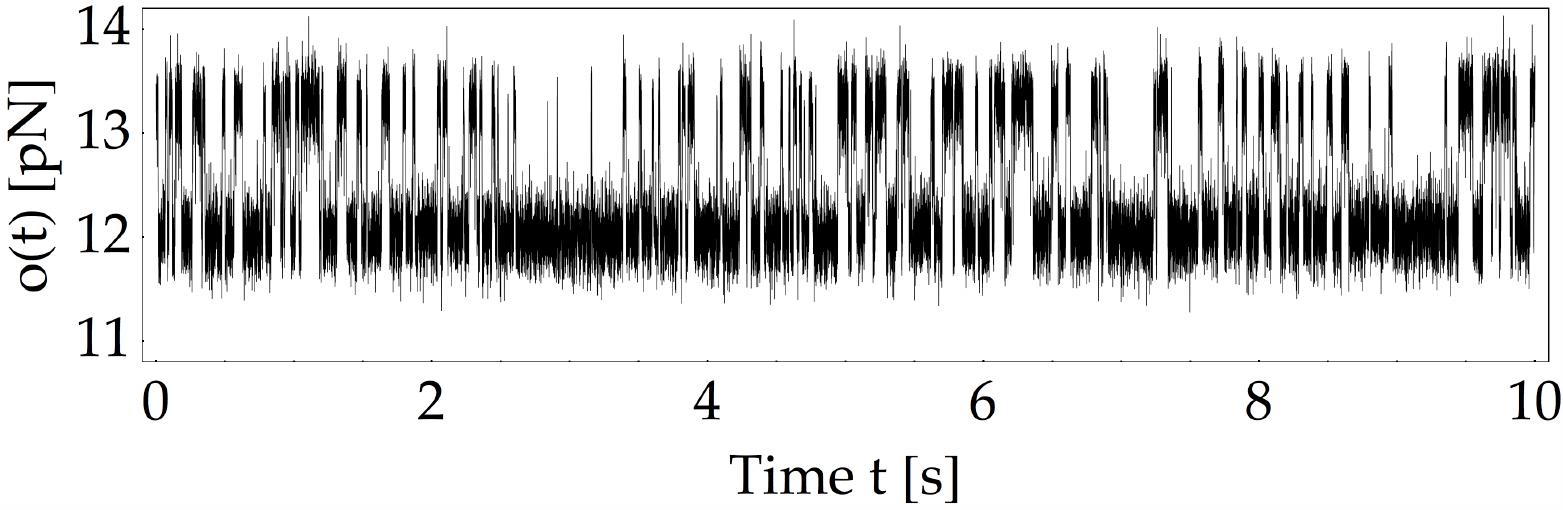}

\includegraphics[width=0.5\columnwidth]{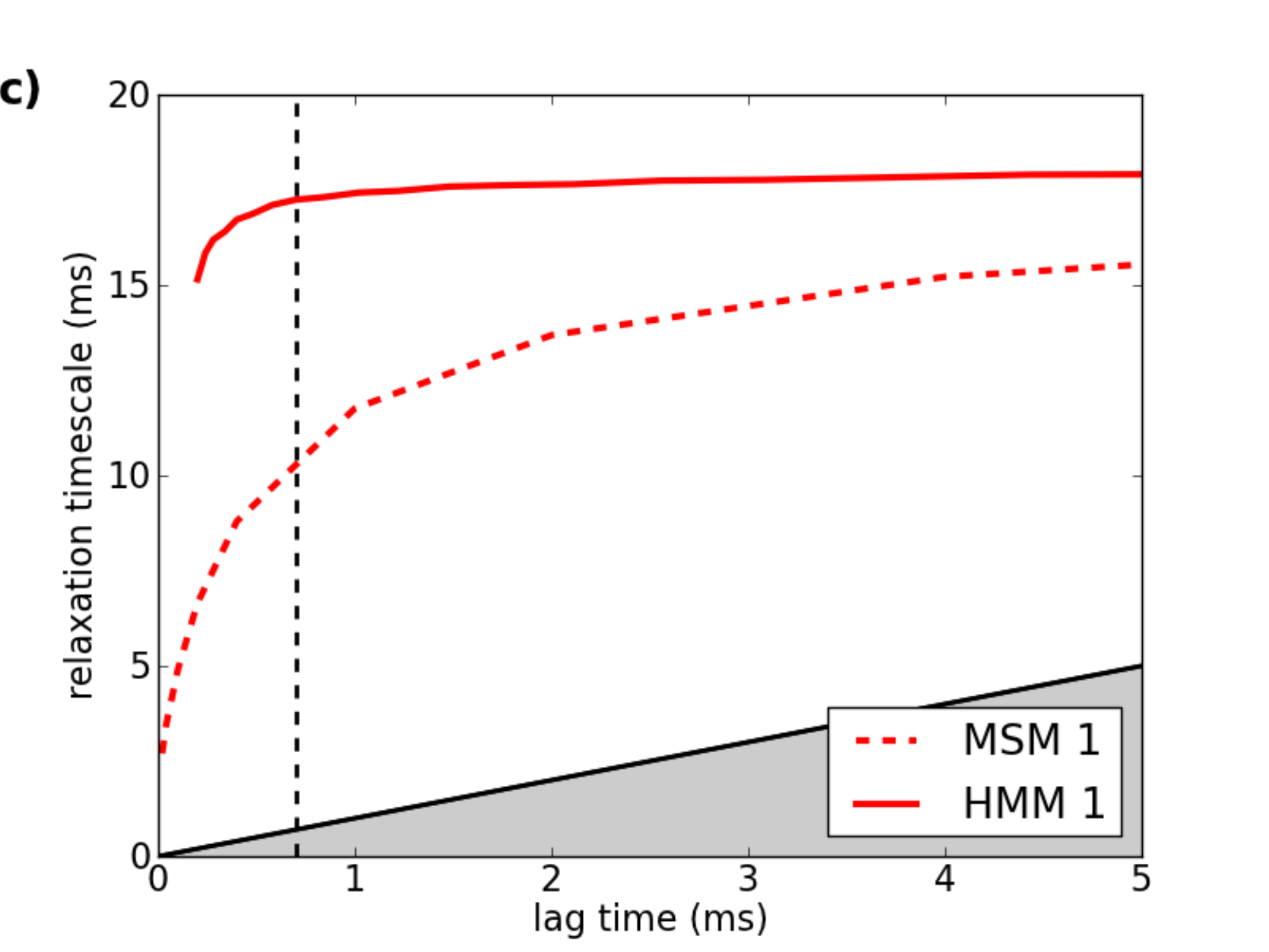}\includegraphics[width=0.5\columnwidth]{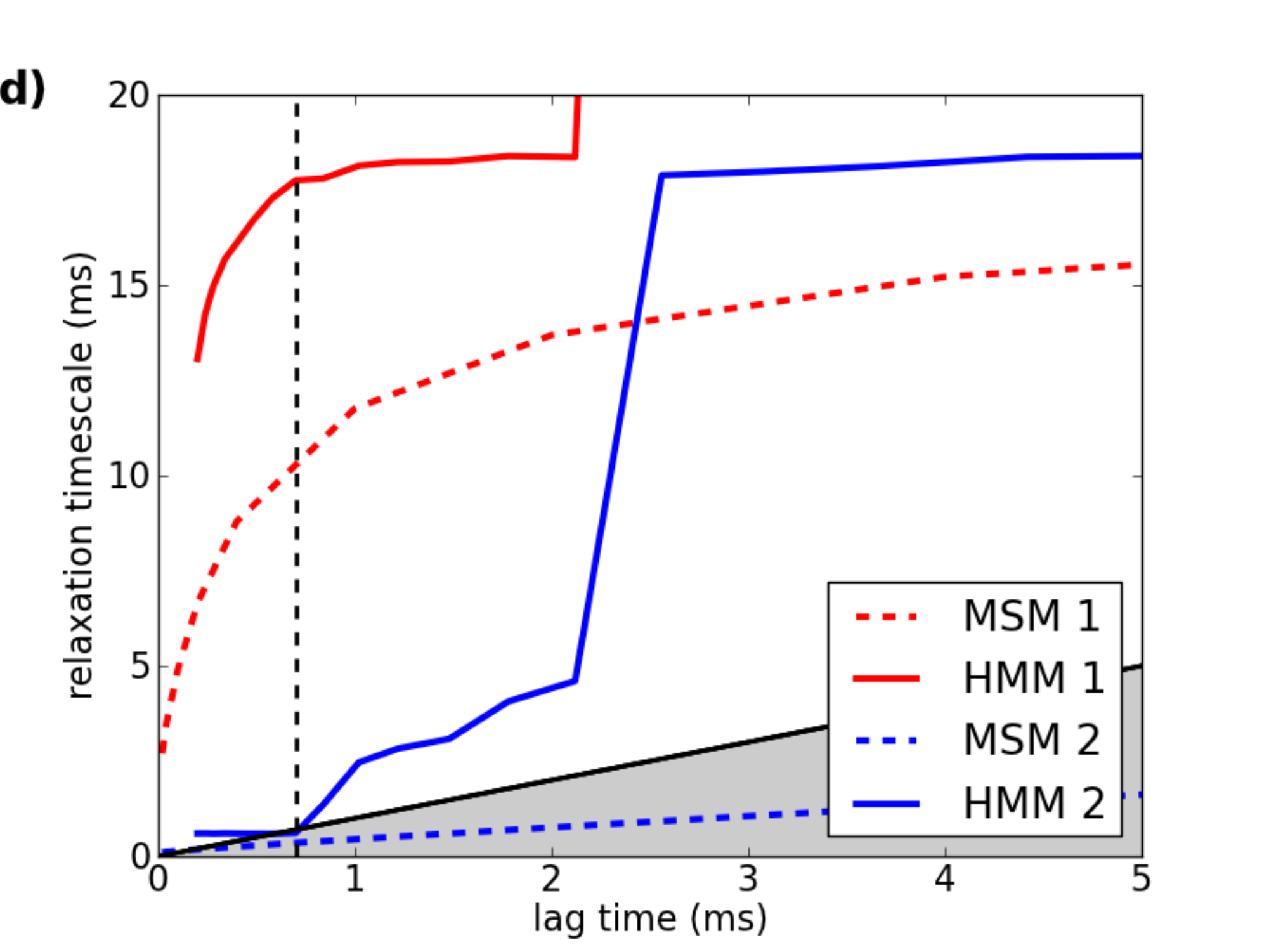}

\includegraphics[width=0.5\columnwidth]{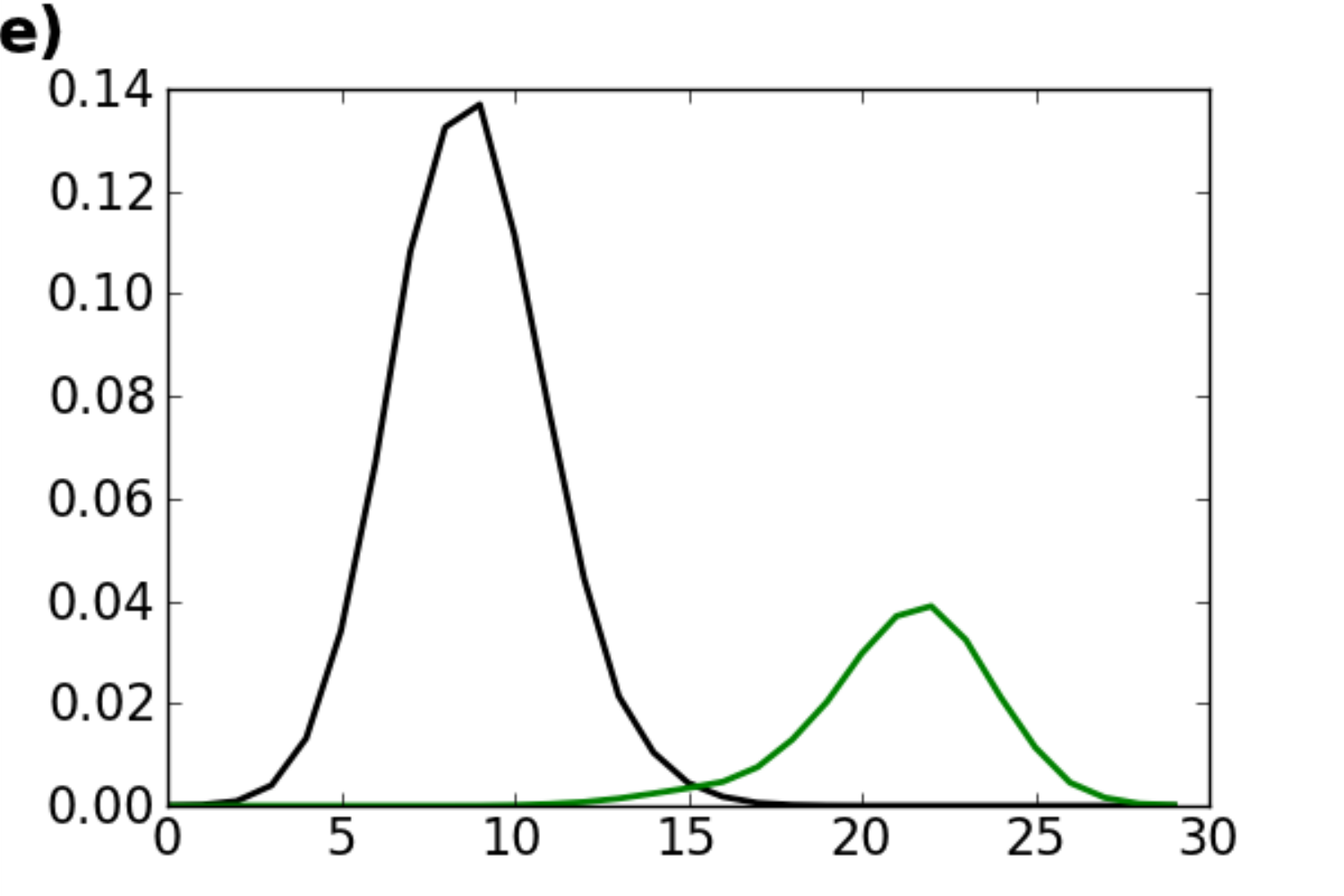}\includegraphics[width=0.5\columnwidth]{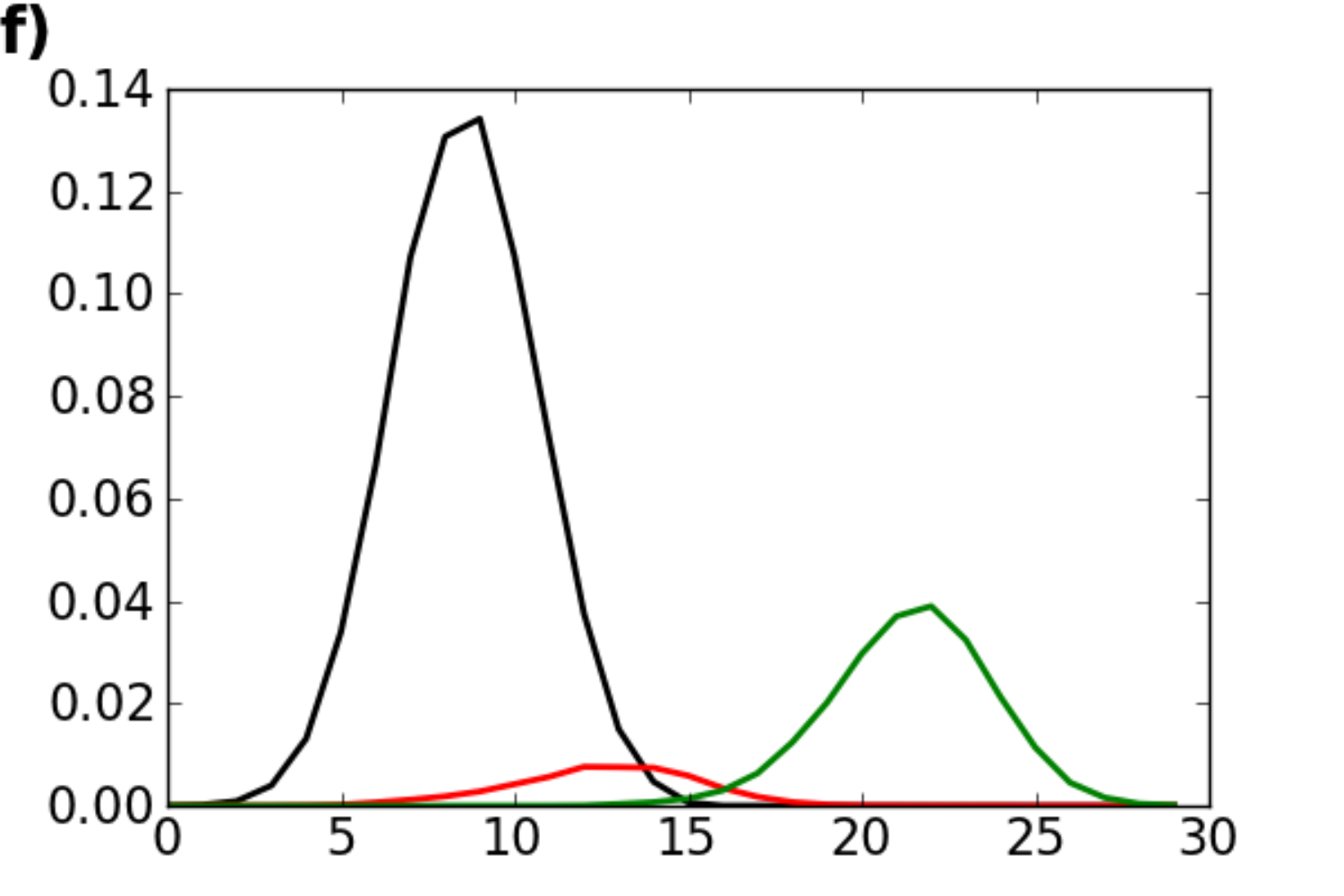}

\includegraphics[width=1\columnwidth]{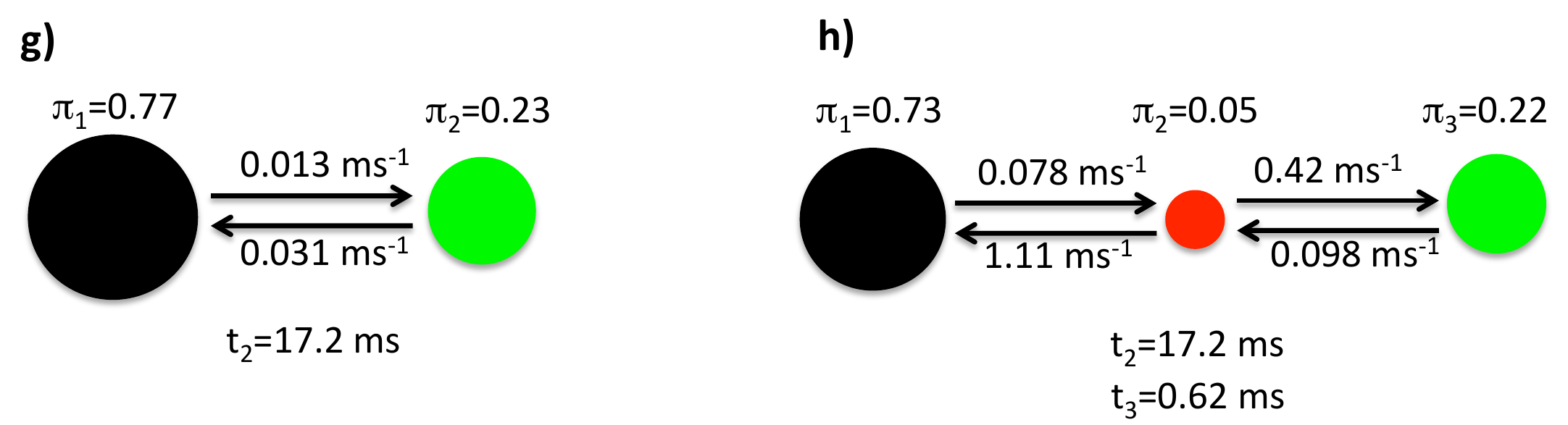}

\caption{\label{fig4_tweezer}Comparison of MSM and HMM for analyzing single-molecule
force-probe data of the RNA hairpin p5ab \cite{elms:2012:biophys-j:force-feedback}.
a) Sketch of the folded secondary structure, b) the optical tweezer
trace. c,d) Relaxation timescales computed by MSMs compared with HMMs
with 2 and 3 states, respectively. e,f) HMM output distributions for
the 2- and 3-state HMM, respectively. g,h) rate matrices of the 2-
and 3-state HMM, respectively.}

\end{figure}

As shown in the second example (Fig. \ref{fig2_3state}), the HMM
estimation can also deal with projections of higher-dimensional dynamical
systems onto low-dimensional observables, provided these projections
do not hide some slow relaxation processes completely. Therefore,
we also illustrate the performance of our method on experimental single-molecule
data. We have chosen optical tweezer measurements of the extension
fluctuations of two biomolecules examined in a recent optical force
spectroscopy study: the p5ab RNA hairpin \cite{elms:2012:biophys-j:force-feedback}.
The p5ab hairpin forms stem-loop structure with a bulge under native
conditions (Fig. \ref{fig4_tweezer}a) and zips/unzips repeatedly
under the conditions used to collect data (Fig. \ref{fig4_tweezer}b).
Experimental force trajectory data were generously provided by the
authors of Ref. \cite{elms:2012:biophys-j:force-feedback}; experimental
details are given therein. The instrument used to collect both datasets
was a dual-beam counter-propagating optical trap. The molecule of
interest was tethered to polystyrene beads by means of dsDNA handles,
with one bead suctioned onto a pipette and the other held in the optical
trap. A piezoactuator controlled the position of the trap and allowed
position resolution to within 0.5 nm, with the instrument operated
in passive (equilibrium) mode such that the trap was stationary relative
to the pipette during data collection. The force on the bead held
in the optical trap was recorded at 50 kHz, with each recorded force
trajectory 60 s in duration. The trajectory shown in Fig. \ref{fig4_tweezer}b
that was chosen for analysis has a relative similar population in
the open and closed states. This experimental construct suffers from
a slow drift in the recorded force trajectory. Although the drift
is very small for the selected trajectory, it may interfere with an
analysis of the slow kinetics; and it will be seen below how the HMM
analysis deals with this.

For the analysis, we discretized the observed force coordinate into
30 regularly spaced bins. Fig. \ref{fig4_tweezer}c,d compares the
performances of 30-state MSMs with two- or three-state HMMs, respectively.
While the MSMs converge only very slowly towards the $\sim17$ ms
timescale associated with the exchange of open and closed states,
both HMMs estimate this timescale robustly after a lag time of $0.7$
ms. Interestingly, for the three-state HMM, there is a switch of eigenvectors
at lag times of 2 to 2.5 ms. While the open/close transition is now
estimated as the second-slowest timescale, the slowest timescale vastly
increases to a timescale on the order of the entire trajectory length.
Inspection of the corresponding eigenvector has confirmed that the
process found by this second timescale indeed corresponds to a slight
shift of the output distributions that captures the small drift that
is present in the trajectory and is associated to a slight shift of
the open and closed force distributions between the beginning and
the end of the trajectory. 

Fig. \ref{fig4_tweezer}d shows that the three-state HMM also finds
a faster process of less than 1 ms for short lag times. Clearly, this
fast process disappears at long lag times, and therefore the blue
curve in Fig. \ref{fig4_tweezer}d leaves this initial plateau after
$\tau>0.7$ ms. However, at $\tau=0.7$ ms both processes are present
in the data, and a three-state HMM can be successfully constructed.
Fig. \ref{fig4_tweezer}e,f show the corresponding HMM output distributions
$\{\boldsymbol{\chi}_{1},\boldsymbol{\chi}_{2},\boldsymbol{\chi}_{3}\}$,
weighted by the stationary probabilities $\{\widetilde{\pi}_{1},\widetilde{\pi}_{2},\widetilde{\pi}_{3}\}$,
thus illustrating where the two or three metastable states are located
in the force coordinate. As expected, the most stable black state
(small forces) and the less stable green state (higher forces) correspond
to the open and closed states of the hairpin. Interestingly, the three-state
HMM identifies a third (red) state that lies ``in between'' open
and closed. This state has so far not been reported. The rate matrix
and stationary probabilities shown in Fig. \ref{fig4_tweezer}h, reveal
that the three states are linearly connected, and the low-populated
red state is a transition state. This rate matrix also indicates that
the intermediate state has a lifetime of about 0.65 ms.

\section{Conclusions}

We have introduced the concept of projected Markov models (PMMs) and
established a connection between conformation dynamics, PMMs, hidden
Markov models (HMMs) and the widely used Markov state models (MSMs).
When observing the continuous and full-phase space dynamics on some
(possibly coarse) set of discrete clusters, the true kinetics is described
by PMMs, rather than with MSMs, although MSMs are very widely used
on this discrete dynamics. MSMs are therefore just an approximation
of the discrete dynamics, which are not actually Markovian. 

Currently, no efficient approach for directly estimating PMMs is available.
Here, we have shown that in the important setting where the dynamics
are metastable, with rarely populated transition regions, and there
is a timescale separation after the first $m-1$ slow relaxation timescales
($m$ could be any number, but is usually small), PMMs can be approximated
with HMMs. This is an important result, because HMMs can be efficiently
estimated by maximum-likelihood or Bayesian techniques, and will in
this setting give the correct estimate of the slow molecular kinetics
- without the systematic bias induced by the Markovianity assumption
of MSMs. 

HMMs are then estimated with an $m\times m$ transition matrix describing
the dynamics between the $m$ hidden states, and each hidden state
associated with an $n$-element probability vector containing the
probability that the hidden state will appear in one of the $n$ discrete
clusters. In order to successfully and reliably conduct the HMM estimation
for large values of $n$, it is important to have a good starting
guess of the transition matrix and the output probability matrix.
Here, we have also made a new connection between MSMs and HMMs and
shown that the initial HMM transition matrix and output probability
matrix can be computed from a set of established algebraic transformations
of an MSM transition matrix. 

We have shown that a vast number of relevant thermodynamics, kinetic,
and mechanistic quantities that are commonly computed from MSMs can
also be computed from HMMs. Notably, this includes kinetic experimental
observables such as time-correlation functions and time-dependent
expectation values of triggered dynamics. These experimentally observable
quantities occur in a functional form that can be readily interpreted
by assigning experimentally-measurable relaxation timescales to the
experimentally not directly measurable structural changes.

Thereby, PMMs, and their HMM-approximations, are invoked as a new
modeling framework for slow molecular kinetics, and a real alternative
to MSMs. Future studies will extend this framework, e.g. by addressing
the computation of the statistical error of the present HMMs via a
full Bayesian analysis.

\section*{Acknowlegdements}

We are grateful to the authors of \cite{elms:2012:biophys-j:force-feedback}
for providing us with the force-probe optical tweezer data and D.E.
Shaw research (especially the authors of \cite{Shaw_Science10_Anton})
for publishing the 1 ms BPTI trajectory. We would also like to thank
Gerhard Hummer (MPI Frankfurt, Germany) for having asked the question
whether a PMM isn't exactly the same like a HMM, during the Cecam
protein folding meeting in Zürich 2012 --- this has been one of the
triggers for the present study. We are grateful for the continuous
support of our work by Christof Schütte (FU Berlin). This work was
funded by Deutsche Forschungsgemeinschaft grants WU 744/1-1, the international
postdoctoral fellow program of the Einstein foundation Berlin, and
ERC starting grant ``pcCell''.

\section*{Appendix A: Proof that an observed metastable Markov process with
$m$ slow relaxation processes is equivalent to a $m$-state HMM }

We consider the dynamics of the Markov process $\mathbf{z}_{t}\in\Omega$,
in the full-dimensional phase space. In this section we do \emph{not
yet} consider any projection to an observation space. The purpose
of this section is to investigate if the following two processes are
equivalent:

\emph{Definition:}\textbf{\emph{ $m$-timescale Markov}}\emph{ }\textbf{\emph{process}}\emph{:
A reversible and ergodic Markov process with $m$ dominant slow processes.
We assume that we work at a lag time $\tau$, at which the all other
processes have decayed. Thus the spectrum is assumed to be $1,\lambda_{2}...,\lambda_{m},0,...,0$).}

\emph{Definition:}\textbf{\emph{ $m$-state hybrid}}\emph{ }\textbf{\emph{process}}\emph{:
A $m\times m$ Markov chain where each state has a fixed output distribution
$\rho_{k}(\mathbf{z})$, $\mathbf{z}\in\Omega$. The process consists
of propagating the Markov chain in time. At every time instant, we
draw an independent random number from $\rho_{k}(\mathbf{z})$ where
$k$ is the current discrete state.}

These two are equivalent if their transition kernels are identical:
\[
p_{\tau}^{(1)}(\mathbf{z}_{0},\mathbf{z}_{\tau})=p_{\tau}^{(2)}(\mathbf{z}_{0},\mathbf{z}_{\tau})
\]
or, equivalently, if their correlation densities are identical:
\begin{eqnarray*}
\mu^{(1)}(\mathbf{z}_{0})p_{\tau}^{(1)}(\mathbf{z}_{0},\mathbf{z}_{\tau}) & = & \mu^{(2)}(\mathbf{z}_{0})p_{\tau}^{(2)}(\mathbf{z}_{0},\mathbf{z}_{\tau})\\
c_{\tau}^{(1)}(\mathbf{z}_{0},\mathbf{z}_{\tau}) & = & c_{\tau}^{(2)}(\mathbf{z}_{0},\mathbf{z}_{\tau})
\end{eqnarray*}

where $\mu$ is the stationary distribution of the respective process.
We write down the corresponding correlation densities:
\begin{enumerate}
\item \textbf{$m$-timescale Markov} \textbf{process}: see \cite{NoeNueske_MMS13_VariationalApproach}
\begin{eqnarray}
c_{\tau}(\mathbf{z}_{0},\mathbf{z}_{\tau}) & = & \mu(\mathbf{z}_{0})\mu(\mathbf{z}_{\tau})+\sum_{k=2}^{m}\mathrm{e}^{-\kappa_{k}\tau}\phi_{k}(\mathbf{z}_{0})\phi_{k}(\mathbf{z}_{\tau})\label{eq_c(x,y)_mMetastable}
\end{eqnarray}

\item \textbf{$m$-state hybrid} \textbf{process}: We use the $m\times m$
transition matrix $\widetilde{\mathbf{T}}(\tau)$ that is reversible
with respect to its stationary distribution $\widetilde{\boldsymbol{\pi}}$,
and the corresponding correlation matrix $\widetilde{\mathbf{C}}(\tau)=\widetilde{\boldsymbol{\Pi}}\widetilde{\mathbf{T}}(\tau)$.
At every time step, the process generates output by drawing independent
random variables from the continuous output functions $\rho_{i}$
associated to the current state $i$:
\begin{eqnarray}
c_{\tau}(\mathbf{z}_{0},\mathbf{z}_{\tau}) & = & \sum_{i,j}\rho_{i}(\mathbf{z}_{0})\widetilde{c}_{ij}(\tau)\rho_{j}(\mathbf{z}_{\tau})\nonumber \\
 & = & \sum_{k}\sum_{i,j}\rho_{i}(\mathbf{z}_{0})\tilde{l}_{ki}\lambda_{k}(\tau)\tilde{l}_{kj}\rho_{j}(\mathbf{z}_{\tau})\nonumber \\
 & = & \sum_{i,j}\pi_{i}\rho_{i}(\mathbf{z}_{0})\pi_{j}\rho_{j}(\mathbf{z}_{\tau})\label{eq_c(x,y)_mHybrid}\\
 &  & +\sum_{k}\mathrm{e}^{-\kappa_{k}\tau}\sum_{i,j}\tilde{l}_{ki}\rho_{i}(\mathbf{z}_{0})\tilde{l}_{kj}\rho_{j}(\mathbf{z}_{\tau})
\end{eqnarray}

\end{enumerate}
In order to show (1)$\equiv$(2), we must show that the expansion
of eigenfunctions into a basis of state output functions 
\begin{equation}
\phi_{k}(\mathbf{z})=\sum_{i}\tilde{l}_{ki}\rho_{i}(\mathbf{z})\label{eq_eigenfunction-representation}
\end{equation}
is feasible. We immediately see that this implies a \textbf{necessary
condition}: in the expansion above, the normalization conditions of
eigenfunctions imply:
\begin{eqnarray}
\langle\phi_{k}\mid\phi_{o}\rangle_{\mu^{-1}} & = & \langle\sum_{i}\tilde{l}_{ki}\rho_{i}(\mathbf{z})\mid\sum_{j}\tilde{l}_{oj}\rho_{j}(\mathbf{z})\rangle_{\mu^{-1}}\nonumber \\
 & = & \sum_{i,j}\tilde{l}_{ki}\tilde{l}_{oj}\langle\rho_{i}\mid\rho_{j}\rangle_{\mu^{-1}}\nonumber \\
 & = & \sum_{i,j}\tilde{l}_{ki}\tilde{l}_{oj}s_{ij}\nonumber \\
 & = & \tilde{\mathbf{l}}_{k}^{T}\mathbf{S}\tilde{\mathbf{l}}_{o}\label{eq_nonoverlap_condition}
\end{eqnarray}
where 
\[
s_{ij}:=\langle\rho_{i}\mid\rho_{j}\rangle_{\mu^{-1}}
\]

is the overlap matrix of basis functions. We have to fulfill
\begin{eqnarray*}
\mathbf{L}\mathbf{S}\mathbf{L}^{T} & = & \mathbf{Id}\\
\mathbf{S} & = & \mathbf{L}^{-1}\mathbf{L}^{-T}=(\mathbf{L}^{T}\mathbf{L})^{-1}=(\mathbf{L}^{T}\mathbf{R}^{T}\boldsymbol{\Pi})^{-1}=\boldsymbol{\Pi}^{-1}.
\end{eqnarray*}
but that means that $\mathbf{S}$ has to be a diagonal matrix. Since
the output distributions $\rho_{i}$ are non-negative, $\mathbf{S}$
can only be diagonal if the sets on which the $\rho_{i}$ are non-zero
do not overlap in the full state space. This condition is necessary
for both directions of the proof. 

This observation suggests that the two processes \textbf{$m$-timescale
Markov} \textbf{process} and \textbf{$m$-state hybrid} \textbf{process}
are generally not equivalent, but equivalance is possible when the
nonoverlap condition $\langle\rho_{i}\mid\rho_{j}\rangle=0$ for $i\neq j$
is used as a condition. Additionally, it has been observed that the
weighted eigenfunction $\psi_{i}=\mu^{-1}\phi_{i}$ are approximately
constant on the metastable sets, a property that will be required
later. Therefore, we define a variation of the m-process Markov

\emph{Definition:}\textbf{\emph{ $m$-metastable Markov}}\emph{ }\textbf{\emph{process}}\emph{:
is a }\textbf{\emph{$m$-}}\emph{timescale Markov}\textbf{\emph{ }}\emph{process
with the following additional properties: Let $\{\rho_{i}\}_{i=1}^{m}$
be a set of non-overlapping probability density functions, and let
$\{A_{1},\ldots,A_{m}\}$ be a partition of $\Omega$ defined as
\begin{equation}
\mathbf{z}\in A_{i}\Leftrightarrow\rho_{i}\left(\mathbf{z}\right)>0\label{eq_decomposable-statdist}
\end{equation}
and
\begin{equation}
\mathbf{z}\in A_{i}\Leftrightarrow\frac{\phi_{i}(\mathbf{z})}{\mu(\mathbf{z})}=const.\label{eq_constant-psi}
\end{equation}
}In this definition, the sets $A_{1},...,A_{m}$ are metastable sets
and the boundaries between them are the transition states. The definition
represents an idealized metastable Markov process: The decomposability
of $\mu$ into $m$ distinct modes implies that transition states
between the metastable sets have no probability density: $\mu(\mathbf{z})=0$.
Furthermore, the assumption that the weighted eigenfunctions $\phi_{i}\mu^{-1}$
are constant on the sets $A_{1},...,A_{m}$ is an idealization of
the fact that these eigenfunctions have been observed to be almost
constant on metastable sets \cite{SchuetteFischerHuisingaDeuflhard_JCompPhys151_146}. 

Therefore, no classical physical system can be an \textbf{$m$}-metastable
Markov\textbf{ }process - whenever transitions between the sets $A_{1},...,A_{m}$
are possible, the dynamical process must travel through the transition
regions, and therefore, $\mu(\mathbf{z})$ will not be exactly zero
in these regions. However, a real metastable system may have transition
states that are rarely populated, and thereby approximate the idealized
\textbf{$m$}-metastable Markov\textbf{ }process.

Below, we will show the following:
\begin{enumerate}
\item \emph{$m$-hybrid process $\Rightarrow$ $m$-metastable Markov}\textbf{\emph{
}}\emph{process}
\item \emph{$m$-metastable Markov}\textbf{\emph{ }}\emph{process $\Rightarrow$
$m$-hybrid process}
\item $m$-metastable PMM $\Leftrightarrow$ $m$-state HMM
\end{enumerate}
From 1 and 2 it is obvious that \emph{$m$-metastable Markov}\textbf{\emph{
}}\emph{process $\Leftrightarrow$ $m$-hybrid process.} The third
step follows from a projection of these full phase-space processes
on the observed discrete clusters.

\subsection*{A.1 A $m$-hybrid process is a $m$-metastable Markov process}

Suppose that the stochastic process $\{\mathbf{z}_{t}\}$ in the state
space $\Omega$ is generated by a $m$-hybrid dynamics with transition
matrix $\widetilde{\mathbf{T}}\left(\tau\right)$ and output distribution
functions $\{\rho_{i}\}_{i=1}^{m}$, where $\widetilde{\mathbf{T}}\left(\tau\right)$
is reversible with respect to its invariant distribution $\widetilde{\boldsymbol{\pi}}$
which can be decomposed as
\begin{equation}
\widetilde{\mathbf{T}}\left(\tau\right)=\widetilde{\boldsymbol{\Pi}}^{-1}\sum_{i=1}^{m}e^{-\kappa_{i}\tau}\tilde{\mathbf{l}}_{i}\tilde{\mathbf{l}}_{i}^{\mathrm{T}}\label{eq:hmm-deomposition}
\end{equation}
where $\widetilde{\boldsymbol{\Pi}}=\mathrm{diag}\left(\boldsymbol{\pi}\right)$,
$\tilde{\mathbf{l}}_{i}$ denotes the $i$-th left eigenvector of
$\widetilde{\mathbf{T}}\left(\tau\right)$. The eigenvectors are normalized
such that they satisfy $\tilde{\mathbf{l}}_{1}=\widetilde{\boldsymbol{\pi}}$
and $\tilde{\mathbf{l}}_{i}^{\mathrm{T}}\widetilde{\boldsymbol{\Pi}}^{-1}\tilde{\mathbf{l}}_{j}=\delta_{ij}$.
We now prove that $\{\mathbf{z}(t)\}$ is also a $m$-metastable Markov
process. Note that for any $t_{1}>t_{2}\ge0$ and $B\subset\Omega$,
\begin{eqnarray}
\mathbb{P}\left(\mathbf{z}_{t_{1}}\in B\mid\{\mathbf{z}_{t}\}_{t=0}^{t_{2}}\right) & = & \sum_{i}\mathbb{P}\left(\mathbf{z}_{t_{1}}\in B,s_{t_{2}}=i\mid\{\mathbf{z}_{t}\}_{t=0}^{t_{2}}\right)\nonumber \\
 & = & \sum_{i}\mathbb{P}\left(\mathbf{z}_{t_{1}}\in B\mid s_{t_{2}}=i,\{\mathbf{z}_{t}\}_{t=0}^{t_{2}}\right)\nonumber \\
 &  & \times\mathbb{P}\left(s\left(t_{2}\right)=i\mid\{\mathbf{z}_{t}\}_{t=0}^{t_{2}}\right)\\
 & = & \sum_{i}\mathbb{P}\left(\mathbf{z}_{t_{1}}\in B\mid s_{t_{2}}=i\right)\nonumber \\
 &  & \times\mathbb{P}\left(s_{t_{2}}=i\mid\mathbf{z}_{t_{2}}\right)\\
 & = & \mathbb{P}\left(\mathbf{z}_{t_{1}}\in B\mid\mathbf{z}_{t_{2}}\right)
\end{eqnarray}
where $s_{t}$ denotes the state of the HMM at time $t$. Therefore,
$\{\mathbf{z}_{t}\}$ is a Markov process.

Furthermore we have the correlation density given in (\ref{eq_c(x,y)_mHybrid})
with the eigenfunction representation (\ref{eq_eigenfunction-representation}).
Using the non-overlap condition (\ref{eq_nonoverlap_condition}),
these eigenfunctions have the correct normalization: 
\begin{eqnarray*}
\left\langle \phi_{i}\mid\phi_{j}\right\rangle _{\mu^{-1}} & = & \mathbf{l}_{i}^{T}\mathbf{S}\mathbf{l}_{j}=\delta_{ij}
\end{eqnarray*}
Therefore $\{\mathbf{z}_{t}\}$ is a $m$-metastable Markov process.

\subsection*{A.2 A $m$-metastable Markov process is a $m$-hybrid process}

Suppose that the stochastic process $\{\mathbf{z}_{t}\}$ is a $m$-metastable
Markov process in state space $\Omega$. Then its correlation density
is given by (\ref{eq_c(x,y)_mMetastable}) and the propagator eigenfunctions
$\{\phi_{i}\}_{i=1}^{m}$, where $\phi_{1}=\mu$ is the stationary
distribution of $\{\mathbf{z}(t)\}$, satisfy the orthogonality conditions
$\left\langle \phi_{i}\mid\phi_{j}\right\rangle _{\mu^{-1}}=\delta_{ij}$.
From (\ref{eq_decomposable-statdist}), we can directly follow
\[
\langle\rho_{i}\mid\rho_{j}\rangle_{\mu^{-1}}=0\:\:\:\:\forall i\neq j
\]
and thus, every density can be described as a linear combination of
basis functions:
\begin{eqnarray*}
\mu(\mathbf{z}) & \in & \mathrm{span}\left(\rho_{1},\ldots,\rho_{m}\right)
\end{eqnarray*}
Combining this result with (\ref{eq_constant-psi}), it follows that
the entire set of propagator eigenfunctions must be expressable in
terms of such linear combinations:
\begin{equation}
\phi_{1},\ldots,\phi_{m}\in\mathrm{span}\left(\rho_{1},\ldots,\rho_{m}\right)
\end{equation}
We call the coefficients required to represent the eigenfunctions
$\{\phi_{i}\}_{i=1}^{m}$ in the basis $\{\rho_{j}\}_{j=1}^{m}$,
$\tilde{l}_{ij}$:
\[
\phi_{i}=\sum_{j}\tilde{l}_{ij}\rho_{j}
\]
and define:
\begin{eqnarray}
\widetilde{\boldsymbol{\pi}}=\left[\tilde{\pi}_{i}\right] & := & \tilde{\mathbf{l}}_{1}\\
\widetilde{\mathbf{T}}\left(\tau\right)=\left[\widetilde{T}_{ij}\left(\tau\right)\right] & := & \widetilde{\boldsymbol{\Pi}}^{-1}\cdot\sum_{k=1}^{m}e^{-\kappa_{k}\tau}\tilde{\mathbf{l}}_{k}\tilde{\mathbf{l}}_{k}^{\mathrm{T}}
\end{eqnarray}
with $\widetilde{\boldsymbol{\Pi}}=\mathrm{diag}\left(\widetilde{\boldsymbol{\pi}}\right)$.
From these definitions, it follows that
\begin{eqnarray}
\tilde{\pi}_{i} & = & \int_{A_{i}}\sum_{j}\widetilde{\pi}_{j}\rho_{j}\left(\mathbf{z}\right)\mathrm{d}\mathbf{z}\nonumber \\
 & = & \mathbb{P}\left(\mathbf{z}_{t}\in A_{i}\right)
\end{eqnarray}
and

\begin{eqnarray}
\tilde{\pi}_{i}\tilde{T}_{ij}\left(\tau\right) & = & \sum_{k=1}^{m}\exp\left(-\kappa_{k}\tau\right)\tilde{l}_{ki}\tilde{l}_{kj}\nonumber \\
 & = & \sum_{k=1}^{m}\exp\left(-\kappa_{k}\tau\right)\int_{A_{i}}\tilde{l}_{ki}\rho_{i}\left(\mathbf{z}_{0}\right)\mathrm{d}\mathbf{z}_{0}\nonumber \\
 &  & \times\int_{A_{j}}\tilde{l}_{kj}\rho_{j}\left(\mathbf{z}_{\tau}\right)\mathrm{d}\mathbf{z}_{\tau}\\
 & = & \sum_{k=1}^{m}\exp\left(-\kappa_{k}\tau\right)\int_{A_{i}}\left(\sum_{a}\tilde{l}_{ka}\rho_{a}\left(\mathbf{z}_{0}\right)\right)\mathrm{d}\mathbf{z}_{0}\nonumber \\
 &  & \times\int_{A_{j}}\left(\sum_{b}\tilde{l}_{kb}\rho_{b}\left(\mathbf{z}_{\tau}\right)\right)\mathrm{d}\mathbf{z}_{\tau}\\
 & = & \sum_{k=1}^{m}\exp\left(-\kappa_{k}\tau\right)\int_{A_{i}}\int_{A_{j}}\phi_{k}\left(\mathbf{z}_{0}\right)\phi_{k}\left(\mathbf{z}_{\tau}\right)\mathrm{d}\mathbf{z}_{\tau}\mathrm{d}\mathbf{z}_{0}\nonumber \\
 & = & \int_{A_{i}}\int_{A_{j}}c\left(\mathbf{z}_{0},\mathbf{z}_{\tau}\right)\mathrm{d}\mathbf{z}_{\tau}\mathrm{d}\mathbf{z}_{0}\nonumber \\
 & = & \mathbb{P}\left(\mathbf{z}_{0}\in A_{i},\mathbf{z}_{\tau}\in A_{j}\right)
\end{eqnarray}
Therefore $\widetilde{\boldsymbol{\pi}}$ is a discrete distribution
and $\widetilde{\mathbf{T}}$ is a reversible transition matrix with
respect to $\widetilde{\boldsymbol{\pi}}$, and we can construct a
$m$-state hybrid Markov process with transition matrix $\widetilde{\mathbf{T}}$
and output distributions $\{\rho_{i}\}_{i=1}^{m}$. Noting that
\begin{eqnarray}
\left\langle \tilde{\mathbf{l}}_{i}|\tilde{\mathbf{l}}_{j}\right\rangle _{\tilde{\boldsymbol{\pi}}^{-1}} & = & \int\frac{\left(\sum_{a}\tilde{l}_{ia}\rho_{a}\left(\mathbf{z}\right)\right)\left(\sum_{b}\tilde{l}_{jb}\rho_{b}\left(\mathbf{z}\right)\right)}{\sum_{c}\widetilde{\pi}_{c}\chi_{c}\left(\mathbf{z}\right)}\mathrm{d}\mathbf{z}\nonumber \\
 & = & \left\langle \phi_{i}\mid\phi_{j}\right\rangle _{\mu^{-1}}\nonumber \\
 & = & \delta_{ij}
\end{eqnarray}
and according to the conclusion in the above section, we can conclude
that the dynamics of $\{\mathbf{z}_{t}\}$ can be exactly described
by a $m$-state hybrid Markov process.

\subsection*{A.3 $m$-metastable PMM $\equiv$ $m$-state HMM}

We now consider that the dynamics are observed on a set of $n$ discrete
states $\{S_{1},...,S_{n}\}$. 

It is straightforward to classify the processes after projecting them
onto an observable $y$:
\begin{enumerate}
\item When projecting a $m$-timescale Markov process onto the discrete
partition $\{S_{1},...,S_{n}\}$, we obtain a \textbf{PMM} with $m$
relaxation timescales (\ref{eq_PMM_C}). Therefore, when \textbf{$m$-metastable
Markov process} onto the partition $\{S_{1},...,S_{n}\}$, we also
obtain a PMM. We call this specific PMM a $m$-metastable PMM.
\item When projecting a \textbf{$m$-hybrid process} onto sets $\{S_{i}\}$,
we obtain a $m$-state \textbf{HMM} with $m$ hidden states, the $m\times m$
transition matrix of the $m$-hybrid process as a hidden transition
matrix, and the output probability matrix
\[
\chi_{ki}=\int_{\mathbf{z}\in S_{i}}\rho_{k}(\mathbf{z})\: d\mathbf{z}.
\]

\end{enumerate}
In the sections above we have shown that for the metastable case,
we have the equality

\noindent \begin{center}
\textbf{$m$-metastable Markov dynamics$\:\:\equiv\:\:$$m$-hybrid
dynamics}
\par\end{center}

and thus we have shown

\noindent \begin{center}
\textbf{$m$-metastable PMM}$\equiv$\textbf{$m$-state HMM}
\par\end{center}

\section*{Appendix B. Algorithms and derivations}

\subsection*{B.1 Estimation algorithm}

We summarize by sketching the PMM/HMM estimation algorithm

\begin{algorithm}
\flushleft

\textbf{Input}:

- $N$ trajectories, discretized into $n$ clusters: $S=\{\{s_{t}^{(1)}\},....,\{s_{t}^{(N)}\}\}$

- lag time: $\tau$

- number of slow relaxation processes considered: $m$

\textbf{Algorithm}:
\begin{enumerate}
\item Estimate reversible Markov transition matrix $\mathbf{T}(\tau)\in\mathbb{R}^{n\times n}$
from the discrete trajectory $S$
\item Decompose $\mathbf{T}(\tau)$ into an initial guess for the HMM matrices:
$\boldsymbol{\chi}\in\mathbb{R}^{n\times m}$ and $\widetilde{\mathbf{T}}(\tau)\in\mathbb{R}^{m\times m}$
using PCCA and Equations (\ref{eq_HMM-Chiinit}-\ref{eq_HMM-Tinit}).
\item Optimize $\boldsymbol{\chi}$ and $\widetilde{\mathbf{T}}(\tau)$
using the EM algorithm.
\item Validate model by comparing correlation matrices $\mathbf{C}^{\mathrm{pred}}(\tau)=\boldsymbol{\chi}\widetilde{\boldsymbol{\Pi}}[\widetilde{\mathbf{T}}(\tau_{0})]^{n}\boldsymbol{\chi}^{\top}$
and $\mathbf{C}(\tau)=\boldsymbol{\Pi}\mathbf{T}(\tau)$, or the apparent
relaxation timescales computed from $\mathbf{T}^{\mathrm{pred}}(\tau)=\boldsymbol{\Pi}^{-1}\boldsymbol{\chi}\widetilde{\boldsymbol{\Pi}}[\widetilde{\mathbf{T}}(\tau_{0})]^{n}\boldsymbol{\chi}^{\top}$
and the direct MSM $\mathbf{T}(\tau)$.
\end{enumerate}
\caption{PMM/HMM estimation}
\end{algorithm}

\subsection*{B.2 computing the HMM transition matrix from PCCA memberships}

We use the definition of the coarse-grained transition matrix derived
in \cite{KubeWeber_JCP07_CoarseGraining} 
\[
\widetilde{\mathbf{T}}=(\mathbf{R}\mathbf{I})^{-\top}\mathbf{I}^{\top}\mathbf{P}\mathbf{R}^{\top}
\]
with restriction and interpolation operators:

\begin{eqnarray*}
\mathbf{R} & = & \mathbf{M}^{\top}\\
\mathbf{I} & = & \boldsymbol{\Pi}\mathbf{M}\widetilde{\boldsymbol{\Pi}}^{-1}.
\end{eqnarray*}
By a series of algebraic transformations we obtain:
\begin{eqnarray*}
\widetilde{\mathbf{T}} & = & (\mathbf{R}\mathbf{I})^{-\top}\mathbf{I}^{\top}\mathbf{P}\mathbf{R}^{\top}\\
 & = & (\mathbf{M}^{T}\boldsymbol{\Pi}\mathbf{M}\widetilde{\boldsymbol{\Pi}}^{-1})^{-T}(\boldsymbol{\Pi}\mathbf{M}\widetilde{\boldsymbol{\Pi}}^{-1})^{T}\mathbf{P}\mathbf{M}\\
 & = & \mathbf{M}^{T}\mathbf{P}^{T}\boldsymbol{\Pi}\mathbf{M}\widetilde{\boldsymbol{\Pi}}^{-1}(\mathbf{M}^{T}\boldsymbol{\Pi}\mathbf{M}\widetilde{\boldsymbol{\Pi}}^{-1})^{-1}\\
 & = & \mathbf{M}^{T}\mathbf{C}\mathbf{M}\widetilde{\boldsymbol{\Pi}}^{-1}(\mathbf{M}^{T}\boldsymbol{\Pi}\mathbf{M}\widetilde{\boldsymbol{\Pi}}^{-1})^{-1}\\
 & = & \mathbf{M}^{T}\mathbf{C}\mathbf{M}(\mathbf{M}^{T}\boldsymbol{\Pi}\mathbf{M}\widetilde{\boldsymbol{\Pi}}^{-1}\widetilde{\boldsymbol{\Pi}})^{-1}\\
 & = & \mathbf{M}^{T}\mathbf{C}\mathbf{M}(\mathbf{M}^{T}\boldsymbol{\Pi}\mathbf{M})^{-1}\\
\mathbf{M}\widetilde{\mathbf{T}}\mathbf{M}^{T}\boldsymbol{\Pi}\mathbf{M}\mathbf{M}^{T} & = & \mathbf{M}\mathbf{M}^{T}\mathbf{C}\mathbf{M}\mathbf{M}^{T}\\
\widetilde{\mathbf{T}}\mathbf{M}^{T} & = & (\mathbf{T}\mathbf{M})^{T}\\
\widetilde{\mathbf{T}} & = & \mathbf{M}^{T}\mathbf{T}\mathbf{M}(\mathbf{M}^{T}\mathbf{M})^{-1}
\end{eqnarray*}

\subsection*{B.3 EM implementation}

In order to estimate a discrete HMM using the Baum-Welch EM method,
we iterate the following two steps 
\begin{enumerate}
\item Expectation step: Estimate the hidden path probabilities $\{\boldsymbol{\alpha}_{t}\}^{(k)}$
and $\{\boldsymbol{\beta}_{t}\}^{(k)}$: 
\[
\{\{\boldsymbol{\alpha}_{t}\}^{(k)},\{\boldsymbol{\beta}_{t}\}^{(k)}\}=\arg\max_{\{\boldsymbol{\alpha}_{t}\},\{\boldsymbol{\beta}_{t}\}}\mathbb{P}(\{s_{t}\}\mid\{\boldsymbol{\alpha}_{t}\},\{\boldsymbol{\beta}_{t}\},\tilde{\mathbf{T}}^{(k)},\boldsymbol{\chi}^{(k)})
\]
and compute the log-likelihood $L=\log\mathbb{P}(\{s_{t}\}\mid\tilde{\mathbf{T}},\boldsymbol{\chi})$.
\item Maximization step: Estimate $\widetilde{\mathbf{T}}^{(k+1)}$ and
$\boldsymbol{\chi}^{(k+1)}$.
\[
\{\widetilde{\mathbf{T}}^{(k+1)},\boldsymbol{\chi}^{(k+1)}\}=\arg\max_{\widetilde{\mathbf{T}},\boldsymbol{\chi}}\mathbb{P}(\widetilde{\mathbf{T}},\boldsymbol{\chi}\mid\{s_{t}\},\{\boldsymbol{\alpha}_{t}\}^{(k)},\{\boldsymbol{\beta}_{t}\}^{(k)})
\]

\end{enumerate}
until the increase of the likelihood (\ref{eq_HMM-likelihood}) falls
below a user-defined threshold. While the expectation step is general,
the maximization step must be designed for the specific HMM implementation.
Here, we estimate the quantities $\widetilde{\mathbf{T}},\boldsymbol{\chi}$
as follows: 
\begin{enumerate}
\item From the expectation step, we compute the Baum-Welch count matrix
between hidden states \cite{BaumEtAl_AnnMathStatist70_BaumWelch,Welch_IEEE03_BaumWelchImplementation}:
\begin{eqnarray*}
\widetilde{\mathbf{Z}}_{t} & = & g^{-1}\boldsymbol{\alpha}_{t}\widetilde{\mathbf{T}}^{(k)}\boldsymbol{\beta}_{t+1}\boldsymbol{\chi}_{s_{t}}\\
g & = & \mathbf{1}^{T}\boldsymbol{\alpha}_{t}\widetilde{\mathbf{T}}^{(k)}\boldsymbol{\beta}_{t+1}\boldsymbol{\chi}_{s_{t}}\mathbf{1}
\end{eqnarray*}
where $S$ is a normalization factor ensuring that we only count 1
transition per time step. The total count matrix is given by the sum
over all single-step count matrices:
\[
\widetilde{\mathbf{Z}}=\sum_{t}\widetilde{\mathbf{Z}}_{t}
\]
which may run over multiple trajectories. Given the count matrix $\widetilde{\mathbf{Z}}$,
we estimate the maximum likelihood transition matrix that fulfills
detailed balance using the algorithm described in \cite{PrinzEtAl_JCP10_MSM1}
and implemented in EMMA \cite{SenneSchuetteNoe_JCTC12_EMMA1.2}:
\begin{eqnarray*}
\widetilde{\mathbf{T}} & = & \arg\max_{\widetilde{\mathbf{T}}}\mathbb{P}(\widetilde{\mathbf{Z}}\mid\widetilde{\mathbf{T}})\\
 &  & \mathrm{such}\:\mathrm{that}\:\widetilde{\boldsymbol{\Pi}}\widetilde{\mathbf{T}}=\widetilde{\mathbf{T}}^{\top}\widetilde{\boldsymbol{\Pi}}
\end{eqnarray*}

\item The stationary probability is computed from the hidden transition
matrix
\[
\widetilde{\boldsymbol{\pi}}^{\top}=\widetilde{\boldsymbol{\pi}}^{\top}\widetilde{\mathbf{T}}
\]

\item The output probability distributions are computed by first estimating
histograms:
\[
y_{ij}=y_{ij}^{0}+\sum_{t}\frac{\alpha_{t,i}\beta_{t,i}1(s_{t}=j)}{\sum_{k}\alpha_{t,k}\beta_{t,k}}
\]
where $y_{ij}$ is the estimated number of times that hidden state
$i$ has produced cluster $j$ as an output. $1(s_{t}=j)$ is an indicator
function that is 1 if the cluster trajectory is at state $j$ at time
$t$, and 0 otherwise. $y_{ij}^{0}$ is a prior count, here uniformly
set to $n^{-1}$. Then, the histograms are normalized to:
\[
\chi_{ij}=\frac{y_{ij}}{\sum_{k}y_{kj}}.
\]

\end{enumerate}

\subsection*{B.4 Derivation of experimental observables}

Correlation function between observables $a$ and $b$ for lag time
$\tau=n\tau_{0}$

\begin{eqnarray*}
\mathbb{E}[a(t)b(t+\tau)] & = & \mathbf{a}^{\top}\boldsymbol{\chi}^{\top}\widetilde{\boldsymbol{\Pi}}[\widetilde{\mathbf{T}}(\tau_{0})]^{n}\boldsymbol{\chi}\mathbf{b}\\
 & = & \mathbf{a}^{\top}\boldsymbol{\chi}^{\top}\sum_{i=1}^{m}\lambda_{i}(\tau)\tilde{\mathbf{l}}_{i}\tilde{\mathbf{l}}_{i}^{\top}\boldsymbol{\chi}\mathbf{b}\\
 & = & \sum_{i=1}^{m}\mathrm{e}^{-\tau\kappa_{i}}\langle\mathbf{a},\mathbf{q}_{i}^{l}\rangle\langle\mathbf{b},\mathbf{q}_{i}^{l}\rangle.
\end{eqnarray*}

Relaxation function of observable $a$, starting from the hidden-state
probability distribution $\widetilde{\mathbf{p}}_{0}$:

\begin{eqnarray*}
\mathbb{E}_{\widetilde{\mathbf{p}}_{0}}[a(\tau)] & = & \mathbf{a}^{\top}(\widetilde{\mathbf{p}}_{0}^{\top}[\widetilde{\mathbf{T}}(\tau_{0})]^{n}\boldsymbol{\chi})^{\top}\\
 & = & \mathbf{a}^{\top}\boldsymbol{\chi}^{\top}(\widetilde{\mathbf{p}}_{0}^{\top}[\widetilde{\mathbf{T}}(\tau_{0})]^{n})^{\top}\\
 & = & \mathbf{a}^{\top}\boldsymbol{\chi}^{\top}(\widetilde{\mathbf{p}}_{0}^{\top}\widetilde{\boldsymbol{\Pi}}^{-1}\sum_{i=1}^{m}\lambda_{i}(\tau)\tilde{\mathbf{l}}_{i}\tilde{\mathbf{l}}_{i}^{\top})^{\top}\\
 & = & \sum_{i=1}^{m}\lambda_{i}(\tau)\mathbf{a}^{\top}\boldsymbol{\chi}^{\top}\tilde{\mathbf{l}}_{i}\tilde{\mathbf{l}}_{i}^{\top}\widetilde{\boldsymbol{\Pi}}^{-1}\widetilde{\mathbf{p}}_{0}\\
 & = & \sum_{i=1}^{m}\mathrm{e}^{-\tau\kappa_{i}}\langle\mathbf{a},\mathbf{q}_{i}^{l}\rangle\langle\tilde{\mathbf{l}}_{i}^{\top},\widetilde{\mathbf{p}}_{0}^{*}\rangle.
\end{eqnarray*}
where $\widetilde{\mathbf{p}}{}_{0}^{*}$ is the \emph{excess probability
distribution $\widetilde{\mathbf{p}}{}_{0}^{*}=\Pi^{-1}\widetilde{\mathbf{p}}{}_{0}$.}


\begin{thebibliography}{10}

\bibitem{REVTEX41Control}


\bibitem{apsrev41Control}
08, 1.

\bibitem{BeauchampEtAl_PNAS2012_FoldingMSMs}
K.A. Beauchamp, R.~McGibbon, Y.-S. Lin, and V.S. Pande.
\newblock Simple few-state models reveal hidden complexity in protein folding.
\newblock {\em Proc Natl. Acad. Sci. USA}, 109:17807--17813, 2012.

\bibitem{Bowman_JCP12_BACE}
G.R. Bowman.
\newblock Improved coarse-graining of markov state models via explicit
  consideration of statistical uncertainty.
\newblock {\em J. Chem. Phys.}, 137:134111, 2012.

\bibitem{BowmanEnsignPande_JCTC2010_AdaptiveSampling}
G.R. Bowman, D.L. Ensign, and V.S. Pande.
\newblock {Enhanced Modeling via Network Theory: Adaptive Sampling of Markov
  State Models}.
\newblock {\em J. Chem. Theory Comput.}, 6(3):787--794, March 2010.

\bibitem{BowmanGeissler_PNAS2012_BindingSiteMSM}
G.R. Bowman and P.L. Geissler.
\newblock Equilibrium fluctuations of a single folded protein reveal a
  multitude of potential cryptic allosteric sites.
\newblock {\em Proc. Natl. Acad. Sci. USA}, 109:11681--11686, 2012.

\bibitem{BucheteHummer_JPCB08}
N.V. Buchete and G.~Hummer.
\newblock {Coarse Master Equations for Peptide Folding Dynamics}.
\newblock {\em J. Phys. Chem. B}, 112:6057--6069, 2008.

\bibitem{ChoderaEtAl_JCP07}
J.~D. Chodera, K.~A. Dill, N.~Singhal, V.~S. Pande, W.~C. Swope, and J.~W.
  Pitera.
\newblock {Automatic discovery of metastable states for the construction of
  Markov models of macromolecular conformational dynamics}.
\newblock {\em J. Chem. Phys.}, 126:155101, 2007.

\bibitem{ChoderaEtAl_BiophysJ11_BHMM}
J.D. Chodera, P.~Elms, F.~No{\'e}, B.~Keller, C.M. Kaiser, A.~Ewall-Wice,
  S.~Marqusee, C.~Bustamante, and N.~Singhal Hinrichs.
\newblock Bayesian hidden markov model analysis of single-molecule force
  spectroscopy: Characterizing kinetics under measurement uncertainty.
\newblock {\em http://arxiv.org/abs/1108.1430}, 2011.

\bibitem{DeuflhardWeber_PCCA}
P.~Deuflhard and M.~Weber.
\newblock {Robust Perron cluster analysis in conformation dynamics}.
\newblock {\em ZIB Report}, 03-09, 2003.

\bibitem{Deuflhard_LinAlgAppl_PCCA}
P.~Deulfhard, W.~Huisinga, A.~Fischer, and C.~Sch{\"{u}}tte.
\newblock {Identification of almost invariant aggregates in reversibly nearly
  uncoupled Markov chains}.
\newblock {\em Lin. Alg. Appl.}, 315:39--59, 2000.

\bibitem{DjurdjevacSarichSchuette_MMS10_EigenvalueError}
N.~Djurdjevac, M.~Sarich, and C.~Sch\"{u}tte.
\newblock {Estimating the eigenvalue error of Markov State Models}.
\newblock {\em Multiscale Model. Simul.}, 10:61--81, 2012.

\bibitem{EisenmesserKayKern_Nature2005}
E.Z. Eisenmesser, O.~Millet, W.~Labeikovsky, D.M. Korzhnev, M.~Wolf-Watz, D.A.
  Bosco, J.J. Skalicky, L.E. Kay, and D.~Kern.
\newblock {Intrinsic dynamics of an enzyme underlies catalysis}.
\newblock {\em Nature}, 438(7064):117--121, November 2005.

\bibitem{elms:2012:biophys-j:force-feedback}
P.J. Elms, J.D. Chodera, C.~Bustamante, and S.~Marqusee.
\newblock The limitations of constant-force-feedback experiments.
\newblock {\em Biophys. J.}, 103:1490, 2012.

\bibitem{GansenSeidel_PNAS2009_Nucleosome}
A.~Gansen, A.~Valeri, F.~Hauger, S.~Felekyan, S.~Kalinin, K.~T\'{o}th,
  J.~Langowski, and C.A.M. Seidel.
\newblock {Nucleosome disassembly intermediates characterized by
  single-molecule FRET}.
\newblock {\em Proc. Natl. Acad. Sci. USA}, 106(36):15308--15313, September
  2009.

\bibitem{GebhardRief_PNAS10_AFMEnergyLandscapeProtein}
J.C. Gebhardt, T.~Bornschl\"{o}gl, and M.~Rief.
\newblock {Full distance-resolved folding energy landscape of one single
  protein molecule}.
\newblock {\em Proc. Natl. Acad. Sci. USA}, 107(5):2013--2018, February 2010.

\bibitem{Singhal_JCP07}
N.S. Hinrichs and V.S. Pande.
\newblock {Calculation of the distribution of eigenvalues and eigenvectors in
  Markovian state models for molecular dynamics}.
\newblock {\em J. Chem. Phys.}, 126:244101, 2007.

\bibitem{KellerPrinzNoe_ChemPhysReview11}
B.~Keller, J.-H. Prinz, and F.~No{\'e}.
\newblock Markov models and dynamical fingerprints: Unraveling the complexity
  of molecular kinetics.
\newblock {\em Chem. Phys.}, 396:92--107, 2012.

\bibitem{KubeWeber_JCP07_CoarseGraining}
S.~Kube and M.~Weber.
\newblock {A coarse graining method for the identification of transition rates
  between molecular conformations}.
\newblock {\em J. Chem. Phys.}, 126(2):024103+, 2007.

\bibitem{BaumEtAl_AnnMathStatist70_BaumWelch}
G.~Soules L.~E.~Baum, T.~Petrie and N.~Weiss.
\newblock A maximization technique occurring in the statistical analysis of
  probabilistic functions of markov chains.
\newblock {\em Ann. Math. Statist.}, 41:164--171, 1970.

\bibitem{LindnerEtAl_JCP13_NeutronScatteringI}
B.~Lindner, Z.~Yi, J.-H. Prinz, J.C. Smith, and F.~No\'{e}.
\newblock {Dynamic Neutron Scattering from Conformational Dynamics I: Theory
  and Markov models}.
\newblock {\em J. Chem. Phys}, 2013 (submitted).

\bibitem{LindorffLarsenEtAl_Science11_AntonFolding}
K.~Lindorff-Larsen, S.~Piana, R.O. Dror, and D.E. Shaw.
\newblock How fast-folding proteins fold.
\newblock {\em Science}, 334:517--520, 2011.

\bibitem{MetznerSchuetteVandenEijnden_JCP06_TPT}
P.~Metzner, C.~Sch\"{u}tte, and E.~Vanden-Eijnden.
\newblock {Illustration of transition path theory on a collection of simple
  examples.}
\newblock {\em The Journal of chemical physics}, 125(8), August 2006.

\bibitem{Xie_PRL05_PowerLaw}
W.~Min, G.~Luo, B.~J. Cherayil, S.~C. Kou, and X.~S. Xie.
\newblock {Observation of a Power-Law Memory Kernel for Fluctuations within a
  Single Protein Molecule}.
\newblock {\em Phys. Rev. Lett.}, 94:198302+, 2005.

\bibitem{Molgedey::94}
L.~Molgedey and H.~G. Schuster.
\newblock Separation of a mixture of independent signals using time delayed
  correlations.
\newblock {\em Phys. Rev. Lett.}, 72:3634--3637, 1994.

\bibitem{NeubauerSeidel_JACS2007_Rhodamine}
H.~Neubauer, N.~Gaiko, S.~Berger, J.~Schaffer, C.~Eggeling, J.~Tuma,
  L.~Verdier, C.A.M. Seidel, C.~Griesinger, and A.~Volkmer.
\newblock {Orientational and dynamical heterogeneity of rhodamine 6G terminally
  attached to a DNA helix revealed by NMR and single-molecule fluorescence
  spectroscopy.}
\newblock {\em J. Am. Chem. Soc.}, 129(42):12746--12755, October 2007.

\bibitem{NoeEtAl_PNAS11_Fingerprints}
F.~No{\'e}, S.~Doose, I.~Daidone, M.~L{\"o}llmann, J.D. Chodera, M.~Sauer, and
  J.C. Smith.
\newblock Dynamical fingerprints for probing individual relaxation processes in
  biomolecular dynamics with simulations and kinetic experiments.
\newblock {\em Proc. Natl. Acad. Sci. USA}, 108:4822--4827, 2011.

\bibitem{NoeHorenkeSchutteSmith_JCP07_Metastability}
F.~No{\'e}, I.~Horenko, C.~Sch{\"u}tte, and J.C. Smith.
\newblock {Hierarchical Analysis of Conformational Dynamics in Biomolecules:
  Transition Networks of Metastable States}.
\newblock {\em J. Chem. Phys.}, 126:155102, 2007.

\bibitem{NoeNueske_MMS13_VariationalApproach}
F.~No{\'e} and F.~N{\"u}ske.
\newblock A variational approach to modeling slow processes in stochastic
  dynamical systems.
\newblock {\em SIAM Multiscale Model. Simul.}, 11:635--655, 2013.

\bibitem{NoeSchuetteReichWeikl_PNAS09_TPT}
F.~No{\'e}, C.~Sch{\"u}tte, E.~Vanden-Eijnden, L.~Reich, and T.R. Weikl.
\newblock Constructing the full ensemble of folding pathways from short
  off-equilibrium simulations.
\newblock {\em Proc. Natl. Acad. Sci. USA}, 106:19011--19016, 2009.

\bibitem{PerezEtAl_JCP13_TICA}
G.~Perez-Hernandez, F.~Paul, T.~Giorgino, G.~{de Fabritiis}, and Frank No{\'e}.
\newblock Identification of slow molecular order parameters for markov model
  construction.
\newblock {\em J. Chem. Phys. (in press)}, 2013.

\bibitem{PrinzChoderaNoe_PRL11_RateTheory}
J.-H. Prinz, J.D. Chodera, and F.~No{\'e}.
\newblock Spectral rate theory for two-state kinetics.
\newblock {\em Phys. Rev. X}, 2013 (in revision. Available at:
  http://arxiv.org/abs/1302.6614v1).

\bibitem{PrinzEtAl_JCP10_MSM1}
J.-H. Prinz, H.~Wu, M.~Sarich, B.~Keller, M.~Senne, M.~Held, J.D. Chodera,
  C.~Sch{\"u}tte, and F.~No{\'e}.
\newblock Markov models of molecular kinetics: Generation and validation.
\newblock {\em J. Chem. Phys.}, 134:174105, 2011.

\bibitem{Rabiner_IEEE89_HMM}
L.R. Rabiner.
\newblock {A tutorial on hidden markov models and selected applications in
  speech recognition}.
\newblock In {\em Proceedings of the IEEE}, pages 257--286, 1989.

\bibitem{Roeblitz_PhD}
S.~R\"{o}blitz.
\newblock {\em {Statistical Error Estimation and Grid-free Hierarchical
  Refinement in Conformation Dynamics}}.
\newblock PhD thesis, 2009.

\bibitem{Santoso_PNAS2009_FretPolymerase}
Y.~Santoso, C.M. Joyce, O.~Potapova, L.~Le Reste, J.~Hohlbein, J.P. Torella,
  N.D.F. Grindley, and A.N. Kapanidis.
\newblock {Conformational transitions in DNA polymerase I revealed by
  single-molecule FRET}.
\newblock {\em Proc. Natl. Acad. Sci. USA}, 107(2):715--720, January 2010.

\bibitem{SarichNoeSchuette_MMS09_MSMerror}
M.~Sarich, F.~No{\'e}, and C.~Sch{\"u}tte.
\newblock On the approximation error of markov state models.
\newblock {\em SIAM Multiscale Model. Simul.}, 8:1154--1177, 2010.

\bibitem{SchuetteFischerHuisingaDeuflhard_JCompPhys151_146}
C.~Sch\"{u}tte, A.~Fischer, W.~Huisinga, and P.~Deuflhard.
\newblock {A Direct Approach to Conformational Dynamics based on Hybrid Monte
  Carlo}.
\newblock {\em J. Comput. Phys.}, 151:146--168, 1999.

\bibitem{SchwantesPande_JCTC13_TICA}
C.R. Schwantes and V.S. Pande.
\newblock Improvements in markov state model construction reveal many
  non-native interactions in the folding of ntl9.
\newblock {\em J. Chem. Theory Comput.}, 9:2000--2009, 2013.

\bibitem{SenneSchuetteNoe_JCTC12_EMMA1.2}
M.~Senne, B.~Trendelkamp-Schroer, A.S.J.S. Mey, C.~Sch{\"u}tte, and F.~No{\'e}.
\newblock {EMMA - A software package for Markov model building and analysis}.
\newblock {\em J. Chem. Theory Comput.}, 8:2223--2238, 2012.

\bibitem{Shaw_Science10_Anton}
D.E. Shaw, P.~Maragakis, K.~Lindorff-Larsen, S.~Piana, R.O. Dror, M.P.
  Eastwood, J.A. Bank, J.M. Jumper, J.K. Salmon, Y.~Shan, and W.~Wriggers.
\newblock {Atomic-Level Characterization of the Structural Dynamics of
  Proteins}.
\newblock {\em Science}, 330(6002):341--346, October 2010.

\bibitem{SinghalPande_JCP123_204909}
N.~Singhal and V.~S. Pande.
\newblock {Error analysis and efficient sampling in Markovian state models for
  molecular dynamics}.
\newblock {\em J. Chem. Phys.}, 123:204909, 2005.

\bibitem{SwopeEtAl_JPCB108_6582}
W.~C. Swope, J.~W. Pitera, F.~Suits, M.~Pitman, and M.~Eleftheriou.
\newblock {Describing protein folding kinetics by molecular dynamics
  simulations: 2. Example applications to alanine dipeptide and beta-hairpin
  peptide}.
\newblock {\em Journal of Physical Chemistry B}, 108:6582--6594, 2004.

\bibitem{VoelzPande_JACS10_NTL9}
V.A. Voelz, G.R. Bowman, K.A. Beauchamp, and V.S. Pande.
\newblock {Molecular Simulation of ab Initio Protein Folding for a Millisecond
  Folder NTL9}.
\newblock {\em J. Am. Chem. Soc.}, 132(5):1526--1528, February 2010.

\bibitem{EVandenEijnden_TPT_JStatPhys06}
{W. E} and {E. Vanden-Eijnden}.
\newblock {Towards a Theory of Transition Paths}.
\newblock {\em Journal of Statistical Physics}, 123(3):503--523, May 2006.

\bibitem{Welch_IEEE03_BaumWelchImplementation}
L.R. Welch.
\newblock Hidden markov models and the baum-welch algorithm.
\newblock {\em IEEE Inf. Theory Soc. Newsletter}, 53:1--13, 2003.

\bibitem{WensleyClark_Nature09_FrustrationHelix}
B.G. Wensley, S.~Batey, F.A.C. Bone, Z.M. Chan, N.R. Tumelty, A.~Steward, L.G.
  Kwa, A.~Borgia, A.~Garrido, and J.~Clarke.
\newblock {Experimental evidence for a frustrated energy landscape in a
  three-helix-bundle protein family}.
\newblock {\em Nature}, 463(7281):685--688, February 2010.

\end{thebibliography}
\end{document}